\documentclass[12pt]{iopart}
\usepackage{iopams}
\usepackage{hyperref}

\begin{document}

\title{Non-Markovian dynamics of a biased qubit coupled to a structured bath}

\author{Congjun Gan \footnote{%
\noindent\textit{Fax}: +86-21-5474-1040 \hspace{2pc}
\noindent\textit{Tel.}: +86-21-5474-0674 }, Peihao Huang, Hang Zheng
}

\address{Department of Physics, Shanghai Jiao Tong University,
Shanghai 200240, People's Republic of China}

\ead{gancongjun@sjtu.edu.cn}

\begin{abstract}
A new analytical approach, beyond rotating wave approximation, based
on unitary transformations and the non-Markovian master equation for
the density operator, is applied to treat the biased spin boson
model with a Lorentzian structured bath for arbitrary detunings at
zero temperature. Compared to zero bias, we find that the dynamics
demonstrates two more damping oscillation frequencies and one
additional relaxation frequency for non-zero bias, where one of the
damping oscillation frequencies is a new effect. Analytical
expressions for the non-Markovian dynamics and the corresponding
spectrum, the localized-delocalized transition point, the
coherent-incoherent transition point, the analytical ground energy,
the renormalized tunneling factor and the susceptibility are
determined. The sum rule and the Shiba relation are checked in the
coherent regime.
\end{abstract}



\vspace{2pc} \noindent\textit{Keywords}: Spin boson model,
Lorentzian bath, Non-Markovian

\submitto{\JPCM} 

 \maketitle

\section{Introduction}

\label{sec:intro}

In a fully quantum-mechanical way, the spin boson model (SBM) \cite%
{01Leggett87,02Weiss,03Gri98} is a prominent physical model in the research
of dynamics and decoherence for numerous physical and chemical processes.
Due to its advantage in the quantitative description of quantum bit (qubit),
the SBM has drawn wide interest in the quantum mechanics field. In the last
decade, many promising scalable solid-state qubit schemes have been proposed
and realized \cite{04charge02,05phase02,06aChi03,06bChi03}. Since
controlling decoherence is the dominating strategy in solid-state qubit \cite%
{07decoherence99,08decoherence02}, qubit can be designed to be coupled to a
harmonic oscillator (HO) or detector instead of the dissipative environment
in order to minimize the decoherence. The HO is coupled further to the
environment \cite{06aChi03,06bChi03,09Wallr04,10Thorwart00}, which is
usually characterized by an Ohmic spectral density $J_{\mathrm{Ohm}}(\omega
) $. Such a qubit-HO-environment proposal can be realized as: a flux-qubit
read out by a dc-SQUID \cite{06aChi03,06bChi03,11Wal00} or a qubit placed in
a leaky cavity \cite{06aChi03,06bChi03,10Thorwart00}. Since we usually only
need to consider two primal states in the qubit and the environment is
characterized by the Ohmic bath, as an alternative but equivalent point of
view, such a qubit-HO-environment model can be exactly mapped to the SBM
with a Lorentzian structured bath $J(\omega )$ (see (\ref{J(omega)}) in Sec.~%
\ref{sec:model}) \cite{12Garg85,13LinTian02,14Wal03,15Robertson05}.

Different from the Ohmic bath, the equilibrium dynamics of the SBM with such
a structured bath or the equivalent qubit-HO-environment model is rarely
studied in the papers, such as the studies on a driven qubit coupled to a
structured environment by Grifoni et al. \cite{16Grifoni1,16Grifoni2}.
However, this paper is interested in the regime with static bias and typical
studies include the quasi-adiabatic propagator path integral (QUAPI) \cite%
{17Thorwart04}, the Van Vleck perturbation theory together with a
Born-Markov master equation (VVBM) \cite{18Hausinger08}, the flow equation
renormalization (FER) \cite{19Kleff03,19Wilhelm04,20Kleff04}, the
non-interacting blip approximation (NIBA) \cite{19Wilhelm04,21Nesi07}, an
approximation scheme by introducing a HO displacement operator \cite{22Brito}
and the perturbation method based on unitary transformation \cite{22huang08}%
. It's known that QUAPI is mainly numerical and it is restricted to finite
temperature \cite{20Kleff04}. VVBM works well with finite bias at low
temperature and Ref. \cite{18Hausinger08} is a nice and physically clear
work. However, since VVBM uses the Van Vleck perturbation theory up to the
second order in the qubit-HO coupling to get the eigenvalues and
eigenfunctions of the non-dissipative qubit-HO system and solves a
Born-Markov master equation for the reduced density matrix in the qubit-HO's
eigenbasis, it requires a small qubit-HO coupling and a Born-Markov
approximation \cite{18Hausinger08}. To our knowledge, FER has not studied
the non-equilibrium dynamics and it needs to choose extra setting parameters
for its best results \cite{22Mielke}. Until now, NIBA is not applicable for
non-zero bias at low temperature \cite{18Hausinger08,19Wilhelm04,21Nesi07}.
Ref. \cite{22Brito} presents only results for zero bias near resonance,
meanwhile, Ref. \cite{22huang08} with one unitary transformation only
presents results for zero bias as well.

In this paper, as an extension to Huang and Zheng's work \cite%
{22huang08,22zheng04}, a new analytical approach, beyond rotating wave
approximation (RWA), based on two unitary transformations and the
non-Markovian master equation for the density operator, is applied to treat
the biased spin boson model (SBM) with a Lorentzian structured bath for
arbitrary detunings at zero temperature. One should note that the two
unitary transformations are different from Ref. \cite{22huang08,22zheng04}
and it makes our approach applicable both for non-zero bias and zero bias.
Moreover, within a nontrivial Born approximation\ but without Markovian
approximation, we get the analytical density operator by the master equation
method, which can easily be extended to finite temperature comparing with
Ref. \cite{22zheng04}. Our approach has several advantages. First, both the
localized-delocalized transition point $\alpha _{L}$ and the
coherent-incoherent transition point $\alpha _{c}$ are studied, which have
not been provided so far (except $\alpha _{c}$ with zero bias by Ref. \cite%
{22huang08}). Second, it works well for a wide parameter range: having no
direct restriction on the qubit-HO coupling, both for biased and unbiased,
at arbitrary finite detunings (positive/negative detunings or on resonance)
and with sufficient strong spin-bath coupling as long as $\alpha <\alpha
_{c} $, which is little-studied and beyond the weak coupling regime. Our
results are checked in the exactly solvable special cases. The dynamics and
the corresponding spectrum are compared to the literature results both for
unbiased and biased cases. The Shiba relation and the sum rule have also
been examined.

This paper is organized as follows. Sec.~\ref{sec:model} introduces the
model and our treatment. Meanwhile, the ground state energy, the
renormalized tunneling factor $\eta $ and localized-delocalized transition
point $\alpha _{L}$ are determined. Sec.~\ref{sec:model density_opertator}
presents the master equation and an analytical expression for the density
operator without Markovian approximation. In Sec.~\ref{sec:P(t)}, we have
calculated the non-equilibrium dynamics $P(t)$ and the corresponding
spectrum $S(\omega )$, and presented the physical interpretation. In Sec.~%
\ref{sec:alpha_C}, it shows the susceptibility and the validation of the
Shiba relation. The coherent-incoherent transition point $\alpha _{c}$ is
determined.

\section{Model and treatment}

\label{sec:model}

In a flux qubit system, the qubit is the two macroscopically distinct
quantum states representing clockwise and anticlockwise rotating
supercurrents. The qubit is entangled with the detecting field, which is
itself coupled with the outside noncoherent environment. The qubit can be
characterized by Pauli matrices, the detecting equipment, which is actually
a LC resonant circuit\cite{06aChi03,06bChi03}, can be described by a
harmonic oscillator with a characteristic frequency $\Omega $ and the
environment can be described by a set of harmonic oscillators. Therefore,
the qubit-HO-environment Hamiltonian can be written as (with Planck units $%
\hbar =k_{B}=1$):
\begin{eqnarray}
H &=&-\frac{\Delta }{2}\sigma _{x}+\frac{\epsilon }{2}\sigma _{z}+{\Omega }%
A^{\dagger }A+\sum_{k}\omega _{k}a^{\dagger }a  \nonumber \\
&+&(A^{\dagger }+A)\left[ g\sigma _{z}+\sum_{k}\kappa _{k}(a_{k}^{\dagger
}+a_{k})\right] +(A^{\dagger }+A)^{2}\sum_{k}\frac{\kappa _{k}^{2}}{\omega
_{k}},  \label{Hamiltonian.equalView}
\end{eqnarray}%
where $\Delta $ is the energy difference of the qubit and $\epsilon $ is the
applied bias; $A$ (or $A^{\dag }$) and $a_{k}$ (or $a_{k}^{\dag }$) are the
annihilation (or creation) operators of harmonic oscillators with
frequencies $\Omega $ and $\omega _{k}$'s, respectively; $g$ is the qubit-HO
coupling and $\kappa _{k}$ is the HO-environment coupling relating to the $k$%
th oscillator. The environment is described by the Ohmic spectral density as
$J_{\mathrm{Ohm}}(\omega )\equiv \sum_{k}\kappa _{k}^{2}{\delta }(\omega
-\omega _{k})=\Gamma \omega $, where $\Gamma $ is the dimensionless coupling
constant to describe the Ohmic bath.

As an alternative but equivalent point of view, such a qubit-HO-environment
model in (\ref{Hamiltonian.equalView}) can be exactly mapped to the
conventional SBM with a Lorentzian structured bath. The Hamiltonian reads
\cite{12Garg85,13LinTian02,14Wal03,15Robertson05,22huang08}
\begin{equation}
H=H_{\mathrm{S}}+H_{\mathrm{B}}+H_{\mathrm{I}},  \label{Hamiltonian}
\end{equation}%
where the subscript `$\mathrm{S}$' denotes the spin system, the subscript `$%
\mathrm{B}$' denotes the boson environment and the subscript `$\mathrm{I}$'
denotes the interaction between the spin and the boson environment, with%
\begin{equation}
H_{\mathrm{S}}=-\frac{\Delta }{2}\sigma _{x}+{\frac{\epsilon }{2}}\sigma
_{z},
\end{equation}%
\begin{equation}
H_{\mathrm{B}}=\sum_{k}{\omega _{k}}b_{k}^{\dagger }b_{k},
\end{equation}%
\begin{equation}
H_{\mathrm{I}}=\frac{1}{2}\sigma _{z}\sum_{k}g_{k}(b_{k}^{\dagger }+b_{k}),
\end{equation}%
where $b_{k}^{\dag }$ ($b_{k}$) is the creation (or annihilation) operator
of the $k$th boson mode with frequency $\omega _{k}$; $\sigma _{x}$ and $%
\sigma _{z}$ are Pauli matrices to describe the spin system; $\epsilon $ is
the bias, $\Delta $ is the bare tunneling and $g_{k}$ is the coupling
between the spin and the boson environment. Notice that $g$ and $g_{k}$ are
two different quantities and there is no correlation between them. The boson
environment of the SBM is described by the Lorentzian structured spectral
density, and it reads \cite{12Garg85,13LinTian02,14Wal03,15Robertson05}
\begin{equation}
J(\omega )=\sum_{k}g_{k}^{2}{\delta }(\omega -\omega _{k})=\frac{2\alpha
\omega \Omega ^{4}}{(\Omega ^{2}-\omega ^{2})^{2}+(2\pi \Gamma \omega \Omega
)^{2}},  \label{J(omega)}
\end{equation}%
in which $\alpha =\lim_{\omega \rightarrow 0}J(\omega )/\left( 2\omega
\right) =8{\Gamma }g^{2}/{\Omega ^{2}}$ is the dimensionless coupling
constant.

Our model starts with the SBM in (\ref{Hamiltonian}). In order to take into
account the correlation between the spin and bosons, a unitary
transformation is applied to $H$ to obtain $H^{\prime }=\exp (S)H\exp (-S)$,
where the generator $S=\sum_{k}\left[ \left( g_{k}/\left( 2\omega
_{k}\right) \right) \right] (b_{k}^{\dag }-b_{k})[\xi _{k}\sigma _{z}+(1-\xi
_{k})\sigma _{0}]$. Here we introduce a constant $\sigma _{0}$ and a $k$%
-dependent function $\xi _{k}$, which will be determined later. We rewrite
the transformed $H^{\prime }=H_{0}^{\prime }+H_{1}^{\prime }+H_{2}^{\prime }$
as%
\begin{eqnarray}
&&H_{0}^{\prime }=-{\frac{1}{2}}\eta \Delta \sigma _{x}+{\frac{1}{2}}%
\epsilon \sigma _{z}+\sum_{k}\omega _{k}b_{k}^{\dag }b_{k}  \nonumber \\
&&-\sum_{k}\frac{g_{k}^{2}}{4\omega _{k}}\xi _{k}(2-\xi _{k})-\sum_{k}\frac{%
g_{k}^{2}}{4\omega _{k}}\sigma _{0}^{2}(1-\xi _{k})^{2},
\end{eqnarray}

\begin{equation}
H_{1}^{\prime }={\frac{1}{2}}\sum_{k}g_{k}(1-\xi _{k})(b_{k}^{\dag
}+b_{k})(\sigma _{z}-\sigma _{0})-{\frac{i\sigma _{y}}{2}}\eta \Delta X,
\end{equation}

\begin{eqnarray}
&&H_{2}^{\prime }=-{\frac{1}{2}}\Delta \sigma _{x}\left( \cosh X-\eta
\right) -{\frac{1}{2}}\Delta i\sigma _{y}\left( \sinh X-\eta X\right)
\nonumber \\
&&-\sum_{k}\frac{g_{k}^{2}}{2\omega _{k}}\sigma _{0}(1-\xi _{k})^{2}(\sigma
_{z}-\sigma _{0}),
\end{eqnarray}%
where $X\equiv {\sum_{k}}\left( g_{k}{\xi _{k}}/\omega _{k}\right) {%
(b_{k}^{\dag }-b_{k})}$, and $\eta $ is the thermodynamic average of $\cosh {%
X}$, as%
\begin{eqnarray}
\eta &=&Z^{-1}\mathrm{Tr}\left[ \exp (-\beta H)\cosh X\right]  \nonumber \\
&=&\exp \left[ -\sum_{k}\frac{g_{k}^{2}}{2\omega _{k}^{2}}\xi _{k}^{2}\coth
\left( \frac{\omega _{k}}{2T}\right) \right] ,  \label{eq.eta.T}
\end{eqnarray}%
with $Z=\mathrm{Tr}[\exp (-\beta H)]$ and $T$ is the temperature.

Since the spin and bosons are decoupled in $H_{0}^{\prime }$, it is exactly
solvable. By a unitary matrix $U=\left(
\begin{array}{cc}
u & v \\
v & -u%
\end{array}%
\right) $, with $u=\sqrt{\left( 1-\epsilon /W\right) /2}$, $v=\sqrt{\left(
1+\epsilon /W\right) /2}$ and $W=\sqrt{\epsilon ^{2}+\eta ^{2}\Delta ^{2}}$,
the diagonalized $\tilde{H}_{0}=U^{\dag }H_{0}^{\prime }U$ reads
\begin{eqnarray}
&&\tilde{H}_{0}=-\frac{1}{2}W\sigma _{z}+\sum_{k}\omega _{k}b_{k}^{\dag
}b_{k}  \nonumber \\
&&-\sum_{k}\frac{g_{k}^{2}}{4\omega _{k}}\xi _{k}(2-\xi _{k})-\sum_{k}\frac{%
g_{k}^{2}}{4\omega _{k}}\sigma _{0}^{2}(1-\xi _{k})^{2}.  \label{eq.transH0}
\end{eqnarray}%
The eigenstate of $\tilde{H}_{0}$ is the direct product $|\{n_{k}\},\pm
\rangle $, where $|\pm \rangle $ are the eigenstates of $\sigma _{z}$ with
eigenvalues $\pm 1$ respectively, and $|\{n_{k}\}\rangle $ are the
eigenstates of bosons with $n_{k}$ phonons for the mode $k$. The ground
state of $\tilde{H}_{0}$ is $|g_{0}\rangle =|\{0_{k}\},+\rangle $ and the
lowest exited states are $|\{0_{k}\},-\rangle $, $|\{1_{k}\},+\rangle $ and $%
|\{1_{k}\},-\rangle $.

Similarly, we make the transformations to get $\tilde{H}_{1}=U^{\dag
}H_{1}^{\prime }U$ and $\tilde{H}_{2}=U^{\dag }H_{2}^{\prime }U$, which are
treated as perturbation and they should be as small as possible. For this
purpose, it's determined as $\sigma _{0}=-\epsilon /W$ and $\xi _{k}=\omega
_{k}/(\omega _{k}+W)$. Thus%
\begin{eqnarray}
&&\tilde{H}_{1}=\frac{1}{2}(1-\sigma _{z})\sum_{k}Q_{k}(b_{k}^{\dag }+b_{k})
\nonumber \\
&&+\frac{1}{2}\sum_{k}V_{k}\left[ b_{k}^{\dag }(\sigma _{x}+i\sigma
_{y})+b_{k}(\sigma _{x}-i\sigma _{y})\right] ,  \label{eq.transH1}
\end{eqnarray}%
where $Q_{k}=g_{k}\left[ \epsilon /(\omega _{k}+W)\right] $ and $V_{k}=g_{k}%
\left[ \eta \Delta /(\omega _{k}+W)\right] $. In the SBM, $g_{k}\ll \Delta $%
. $Q_{k}$ and $V_{k}$ can be viewed as the renormalized spin-bath coupling,
and they are always smaller than $g_{k}$ and even smaller for the high
frequencies. Obviously, $\tilde{H}_{1}|g_{0}\rangle =0$. Under the
eigenbasis of $\tilde{H}_{0}$, $\tilde{H}_{1}$ has only off-diagonal terms\
and in the lowest states, it is $\langle \{0_{k}\},-|\tilde{H}%
_{1}|\{1_{k}\},+\rangle =V_{k}$, $\langle \{0_{k}\},-|\tilde{H}%
_{1}|\{1_{k}\},-\rangle =Q_{k}$ and $\langle \{1_{k}\},-|\tilde{H}%
_{1}|\{1_{k^{\prime }}\},+\rangle =0$. Meanwhile, the terms in $\tilde{H}%
_{2} $ are related to the multi-boson transition and their contributions to
the physical quantities are to the fourth order of $g_{k}$ ($O(g_{k}^{4})$).
These are key points in our approach. The transformed Hamiltonian is
approximated as
\begin{eqnarray}
\tilde{H} &=&\tilde{H}_{0}+\tilde{H}_{1}+\tilde{H}_{2}  \nonumber \\
&\approx &\tilde{H}_{0}+\tilde{H}_{1}  \label{Hamiltonian.transformed}
\end{eqnarray}%
in the following. Meanwhile, the previous treatment is an extension to the
one proposed by Ref. \cite{22zheng04}, while our generator $S$ and the
second unitary transformation are different. However, the $k$-dependent
function $\xi _{k}$ and decomposing the Hamiltonian into three parts are
with the same spirit and they have been discussed detailedly in Ref. \cite%
{22zheng04}.

Thus, the ground energy of $\tilde{H}$ is just the same as that of $\tilde{H}%
_{0}$ and it is determined as
\begin{equation}
E_{g}=-\frac{1}{2}W-\sum_{k}\frac{g_{k}^{2}}{4\omega _{k}}\left[ 1-\left(
\frac{\eta \Delta }{\omega _{k}+W}\right) ^{2}\right] .  \label{eq.15}
\end{equation}%
The Hamiltonian $H$ in (\ref{Hamiltonian}) can be solved exactly in two
limits: one is the weak coupling limit with $E_{g}(\alpha \rightarrow 0)=-%
\sqrt{\Delta ^{2}+\epsilon ^{2}}/2$ and the other is the zero tunneling
limit with $E_{g}(\Delta \rightarrow 0)=-|\epsilon
|/2-\sum_{k}g_{k}^{2}/\left( 4\omega _{k}\right) $. The ground energies in (%
\ref{eq.15}) are the same in both limits.

Up to now, the deduction is independent of any specific spectral density and
it is not restricted to zero temperature. In the following, the treatment is
at zero temperature. As shown in $H_{0}^{\prime }$, $\eta $ is the
renormalized tunneling factor. In the limit of zero temperature, it is
\begin{equation}
\eta =\exp \left[ -\int_{0}^{\infty }\frac{J(\omega )\mathrm{d}\omega }{%
2\left( \omega +W\right) ^{2}}\right] .  \label{eq.eta}
\end{equation}%
The integration in (\ref{eq.eta}) can be done to the end, analytically. In
the case of zero bias ($\epsilon =0$), $\eta $ has the same expression as
Huang's and similarly positive change of the tunneling frequency can be
predicted when $\Delta \sim \Omega $, which fails by adiabatic approach \cite%
{22huang08}. Generally, the renormalized tunneling factor $\eta $ is larger
than $0$, which means that there is an effective tunneling between the two
states of the qubit in realistic situation. If the renormalized tunneling
factor suddenly changes to $0$, the localized-delocalized transition happens
and the qubit will be localized in one of the two states where it is located
before the transition.

Fig. 1 shows numerical results of $\eta $ as a function of $\alpha $. For
larger $\Gamma $ (eg. $0.15,0.3$), one can see that $\eta $ suddenly goes to
zero at the localized-delocalized transition point $\alpha =\alpha _{L}$,
where $\eta =0$ for all $\alpha \geq \alpha _{L}$. While for smaller $\Gamma
$ (eg. $0.01$), $\eta $ gradually goes to zero, and we set the cutting at
the value $\eta =0.0001$, which is small enough.

A phase diagram of the delocalized-localized transition point $\alpha _{L}$
vs. bias $\epsilon $ is plotted in Fig. 2 with different $\Gamma $ (=$%
0.01,0.02,0.05,0.1$) and different detunings $\Delta /\Omega $ (=$0.5,1$).
The area of $\alpha <\alpha _{L}$ is called the \textquotedblleft localized
phase\textquotedblright , and the area of $\alpha >\alpha _{L}$ the
\textquotedblleft delocalized phase\textquotedblright . It shows that $%
\alpha _{L}$ increases with increasing $\epsilon $ and it is almost the same
for different $\Delta ^{\prime }$s. The change of $\alpha _{L}$ is
remarkable for larger $\Gamma $ or smaller $\epsilon $. Therefore, one way
by applying a small bias to the qubit can be used to read out the already
localized qubit state, since it will greatly increases $\alpha _{L}$ and the
qubit will be shifted from the localized state to the delocalized one.

\section{Density operator and master equation}

\label{sec:model density_opertator}

In the Schr\"{o}dinger picture, the density operator of the SBM is denoted
as $\rho _{\mathrm{SB}}(t)$ for the Hamiltonian $H$ in (\ref{Hamiltonian})
and the density operator for $\tilde{H}$ in (\ref{Hamiltonian.transformed})
is $\tilde{\rho}_{\mathrm{SB}}(t)=U^{\dag }\exp (S)\rho _{\mathrm{SB}%
}(t)\exp (-S)U$, where the subscript `$\mathrm{SB}$' denotes the total
spin-boson system. In the following, it will be analyzed in the interaction
picture, denoting by a superscript `$\mathrm{I}$' in the operator. $\tilde{H}%
_{0}$ is treated as the unperturbed part and $\tilde{H}_{1}$ is really a
good perturbed part. Moreover, in the interaction picture, it is assumed
that the density operator for $\tilde{H}$ is $\tilde{\rho}_{\mathrm{SB}}^{%
\mathrm{I}}(t)=\tilde{\rho}_{\mathrm{S}}^{\mathrm{I}}(t)\rho _{\mathrm{B}}$,
where $\tilde{\rho}_{\mathrm{S}}^{\mathrm{I}}(t)=\mathrm{Tr_{B}}\tilde{\rho}%
_{\mathrm{SB}}^{\mathrm{I}}(t)$ is the reduced density operator. Within Born
approximation (only keeping the second order of $\tilde{H}_{1}$), we can
obtain the non-Markovian master equation for the reduced density operator
\begin{equation}
\frac{\mathrm{d}}{\mathrm{d}t}\tilde{\rho}_{\mathrm{S}}^{\mathrm{I}%
}(t)=-\int_{0}^{t}\mathrm{Tr_{B}}[\tilde{H}_{1}(t),[\tilde{H}_{1}(t^{\prime
}),\tilde{\rho}_{\mathrm{S}}^{\mathrm{I}}(t^{\prime })\rho _{\mathrm{B}}]]%
\mathrm{d}t^{\prime },  \label{eq.masterEquation}
\end{equation}%
where $\tilde{H}_{1}(t)$ is denoted as the perturbed part $\tilde{H}_{1}$ in
the interaction picture. Since the renormalized spin-bath coupling $Q_{k}$
and $V_{k}$ in $\tilde{H}_{1}$ are always smaller than $g_{k}$, it makes our
Born approximation nontrivial and more reasonable by comparing with the
common used Born approximation \cite{23Burkard2009,23DiVincenzo2005}, which
directly does the perturbation to the second order of $H\mathrm{_{I}}$ in (%
\ref{Hamiltonian}).

The master equation in (\ref{eq.masterEquation}), without Markovian
approximation, can be done to the end with a Laplace transformation and an
inverse-Laplace transformation. Changing from the interaction picture back
to the Schr\"{o}dinger picture, denoting the reduced density operator in the
Schr\"{o}dinger picture as $\tilde{\rho}_{\mathrm{S}}(t)=\left(
\begin{array}{cc}
\tilde{\rho}_{11}(t) & \tilde{\rho}_{12}(t) \\
\tilde{\rho}_{21}(t) & \tilde{\rho}_{22}(t)%
\end{array}%
\right) $ for $\tilde{H}$, at zero temperature, we obtain%
\begin{equation}
\tilde{\rho}_{21}(t)=\frac{\tilde{\rho}_{21}(0)}{2\pi }\int\limits_{-\infty
}^{\infty }\frac{i\exp (-i\omega t)\mathrm{d}\omega }{\omega -W-\Sigma
(\omega )+i\Gamma (\omega )},  \label{eq.rou21.final}
\end{equation}%
and%
\begin{equation}
\tilde{\rho}_{22}(t)=\frac{\tilde{\rho}_{22}(0)}{2\pi }\int\limits_{-\infty
}^{\infty }\frac{i\exp (-i\omega t)\mathrm{d}\omega }{\omega -\Sigma
^{\prime }(\omega )+i\Gamma ^{\prime }(\omega )}.  \label{eq.rou22.final}
\end{equation}%
Abbreviations are used in (\ref{eq.rou21.final}) and (\ref{eq.rou22.final}),
as
\begin{equation}
\Gamma (\omega )=\gamma (\omega )+\frac{\epsilon ^{2}}{\eta ^{2}\Delta ^{2}}%
\gamma (\omega -W),  \label{eq.Gammaw}
\end{equation}%
\begin{equation}
\Sigma (\omega )=R(\omega )+\frac{\epsilon ^{2}}{\eta ^{2}\Delta ^{2}}%
R(\omega -W),  \label{eq.Sigmaw}
\end{equation}%
and%
\begin{equation}
\Gamma ^{\prime }(\omega )=\gamma (W+\omega )+\gamma (W-\omega ),
\label{eq.GammaPriw}
\end{equation}%
\begin{equation}
\Sigma ^{\prime }(\omega )=R(W+\omega )-R(W-\omega ),  \label{eq.SigmaPriw}
\end{equation}%
where $R(\omega )$ and $\gamma (\omega )$ are real and imaginary parts of $%
\sum_{k}V_{k}^{2}/\left( \omega -i0^{+}-\omega _{k}\right) $ ( $0^{+}$ is a
positive infinitesimal introduced by the inverse-Laplace transformation),%
\begin{equation}
R(\omega )=\sum_{\mathbf{k}}\frac{V_{k}^{2}}{\left( \omega -\omega _{\mathbf{%
k}}\right) }=\int\limits_{0}^{\infty }\frac{\eta ^{2}\Delta ^{2}J(\omega
^{^{\prime }})\mathrm{d}\omega ^{^{\prime }}}{(\omega -\omega ^{^{\prime
}})(\omega ^{^{\prime }}+W)^{2}},  \label{eq.Rw}
\end{equation}%
\begin{equation}
\gamma (\omega )=\pi \sum_{\mathbf{k}}V_{k}^{2}\delta \left( \omega -\omega
_{\mathbf{k}}\right) =\frac{\pi \eta ^{2}\Delta ^{2}J(\omega )}{(\omega
+W)^{2}},  \label{eq.gammaw}
\end{equation}%
respectively. Besides, two other terms in $\tilde{\rho}_{\mathrm{S}}(t)$ are
$\tilde{\rho}_{12}(t)=\left[ \tilde{\rho}_{21}(t)\right] ^{\dag }$ and $%
\tilde{\rho}_{11}(t)=1-\tilde{\rho}_{22}(t)$. Since the specific form of $%
J(\omega )$ is not involved, therefore, an analytical expression of the
reduced density operator $\tilde{\rho}_{\mathrm{S}}(t)$ is offered and it is
independent of any specific spectral density.

We assume the initial density operator at $t=0$ is $\rho _{\mathrm{SB}%
}(0)=\exp (-S)\left\vert +\right\rangle \left\langle +\right\vert \rho _{%
\mathrm{B}}\exp (S)$. Thus, the corresponding initial reduced density
operator for $\tilde{H}$ in (\ref{Hamiltonian.transformed}) is
\begin{equation}
\tilde{\rho}_{\mathrm{S}}(0)=\frac{1}{2}\left(
\begin{array}{cc}
1-\epsilon /W & \eta \Delta /W \\
\eta \Delta /W & 1+\epsilon /W%
\end{array}%
\right) .  \label{eq.rou.reduced.init}
\end{equation}

\section{Non-equilibrium dynamics and the physical interpretation}

\label{sec:P(t)}

For the SBM, it is common to evaluate the non-equilibrium dynamics $P(t)$,
as this is the quantity of interest in the experiments. $P(t)$ is also
called the population difference. Following the unitary transforms, it is
determined as%
\begin{eqnarray}
&&P(t)=\mathrm{Tr}_{\mathrm{S}}\left( \mathrm{Tr}_{\mathrm{B}}\left( {\rho }%
_{\mathrm{SB}}(t)\sigma _{z}\right) \right)  \nonumber \\
&=&\mathrm{Tr}_{\mathrm{S}}\left( \mathrm{Tr}_{\mathrm{B}}\left( \exp \left(
-S\right) U\tilde{\rho}_{\mathrm{SB}}(t)U^{\dag }\exp \left( S\right) \sigma
_{z}\right) \right)  \nonumber \\
&=&\frac{\epsilon }{W}\left[ 2\tilde{\rho}_{22}(t)-1\right] +\frac{2\eta
\Delta }{W}\mathrm{Re}\left[ \tilde{\rho}_{21}(t)\right] .
\label{eq.pt.anyInitCase}
\end{eqnarray}%
Substituting (\ref{eq.rou21.final}) and (\ref{eq.rou22.final}) into (\ref%
{eq.pt.anyInitCase}) with the initial condition in (\ref{eq.rou.reduced.init}%
), the dynamics reads
\begin{eqnarray}
&&P(t)=\frac{2\epsilon }{\pi W}\left( 1+\frac{\epsilon }{W}\right)
\int\limits_{0}^{\infty }\mathrm{d}\omega \frac{\cos (\omega t)\Gamma
^{\prime }(\omega )}{\left[ \omega -\Sigma ^{\prime }(\omega )\right] ^{2}+%
\left[ \Gamma ^{\prime }(\omega )\right] ^{2}}  \nonumber \\
&+&\frac{\eta ^{2}\Delta ^{2}}{\pi W^{2}}\int\limits_{0}^{\infty }\mathrm{d}%
\omega \frac{\cos (\omega t)\Gamma (\omega )}{\left[ \omega -W-\Sigma
(\omega )\right] ^{2}+\left[ \Gamma (\omega )\right] ^{2}}-\frac{\epsilon }{W%
}.  \label{eq.pt.analytic}
\end{eqnarray}%
Thus, we end up at an exact analytical expression of the non-Markovian
dynamics $P(t)$ in (\ref{eq.pt.analytic}). As time goes to infinity, we have
the dynamics at the long time limit $P(t\rightarrow \infty )=-\epsilon /W$.

\subsection{Spectrum of the non-Markovian dynamics}

In order to get insight into the dominant frequencies of $P(t)$, a Fourier
transform is applied to (\ref{eq.pt.analytic}) according to%
\begin{equation}
S(\omega )\equiv \int_{-\infty }^{\infty }\mathrm{d}t\cos {(\omega t)}\,P(t).
\end{equation}%
The spectrum $S(\omega )$ is an even function and for $\omega \geq 0$, it is
written as
\begin{eqnarray}
&&S(\omega )=\frac{2\epsilon }{W}\left( 1+\frac{\epsilon }{W}\right) \frac{%
\Gamma ^{\prime }(\omega )}{\left[ \omega -\Sigma ^{\prime }(\omega )\right]
^{2}+\left[ \Gamma ^{\prime }(\omega )\right] ^{2}}  \nonumber \\
&+&\frac{\eta ^{2}\Delta ^{2}}{W^{2}}\frac{\Gamma (\omega )}{\left[ \omega
-W-\Sigma (\omega )\right] ^{2}+\left[ \Gamma (\omega )\right] ^{2}}-\frac{%
2\pi \epsilon }{W}\delta ({\omega }).  \label{eq.sw}
\end{eqnarray}

The frequency property of the dynamics $P(t)$ can be analyzed by $S(\omega )$
directly. On one hand, the first two terms in (\ref{eq.sw}) are
Lorentzian-like functions. On the other hand, $\gamma (\omega )$ is small
when $g$ is small or $\omega $ is away from $\Omega $, thus $\Gamma (\omega
) $ and $\Gamma ^{\prime }(\omega )$, which are functions related to $\gamma
(\omega )$, are usually small. Therefore, the dominant frequencies of $%
S(\omega )$ should be the solutions $\omega _{p}$ of the equation
\begin{equation}
\omega -W-\Sigma (\omega )=0,  \label{eq.omega_p}
\end{equation}%
and the solutions $\omega _{p^{\prime }}$ of the equation
\begin{equation}
\omega -\Sigma ^{\prime }(\omega )=0.  \label{eq.omega_pPri}
\end{equation}

Since the dissipative environment generally adds shift and width to the
dominant frequencies, we can investigate the physical nature in the limit of
small HO-environment coupling ($\Gamma \rightarrow 0$). Consequently, the
spectral density $J(\omega )$ in (\ref{J(omega)}) goes to $\left(
4g^{2}\Omega /\omega \right) \left[ \delta \left( \omega -\Omega \right)
+\delta \left( \omega +\Omega \right) \right] $ and $R(\omega )$ in (\ref%
{eq.Rw}) goes to $4g^{2}\eta ^{2}\Delta ^{2}/\left[ (\omega -\Omega )(\Omega
+W)^{2}\right] $. Therefore, according to (\ref{eq.omega_p}) and (\ref%
{eq.Sigmaw}), the dominant frequencies $\omega _{p}$ are solutions to the
equation%
\begin{equation}
\omega -W=\frac{4g^{2}}{(\Omega +W)^{2}}\left[ \frac{\eta ^{2}\Delta ^{2}}{%
\omega -\Omega }+\frac{\epsilon ^{2}}{\omega -\Omega -W}\right] .
\label{eq.wp}
\end{equation}%
The equation (\ref{eq.wp}) can be solved exactly. If $g^{2}\epsilon ^{2}/%
\left[ \Delta ^{2}(\Omega +W)^{2}\right] \ll 1$, $\omega _{p}$ can be
simplified and approximated as \
\begin{equation}
\omega _{p1,2}\cong \frac{\Omega +W}{2}\pm \sqrt{\left( \frac{\Omega -W}{2}%
\right) ^{2}+\frac{4g^{2}\eta ^{2}\Delta ^{2}}{(\Omega +W)^{2}}}
\label{eq.wp12}
\end{equation}%
where the subscripts `$1,2$' relating to the sign `$+,-$', respectively, and%
\begin{equation}
\omega _{p3}\cong \left( \Omega +W\right) +\frac{W4g^{2}\epsilon ^{2}}{%
\Omega W(\Omega +W)^{2}-4g^{2}\eta ^{2}\Delta ^{2}}.  \label{eq.wp3}
\end{equation}%
Similarly, according to (\ref{eq.omega_pPri}) and (\ref{eq.SigmaPriw}), the
dominant frequencies $\omega _{p^{\prime }}$ are exactly solvable, as%
\begin{equation}
\omega _{p^{\prime }1}=0  \label{eq.wpPri1}
\end{equation}%
and%
\begin{equation}
\omega _{p^{\prime }2,3}=\pm \sqrt{\left( W-\Omega \right) ^{2}+\frac{%
8g^{2}\eta ^{2}\Delta ^{2}}{(\Omega +W)^{2}}}.  \label{eq.wpPri23}
\end{equation}%
Since $S(\omega )$ is an even function, we only need to consider the
non-negative part ($\omega \geq 0$). Thus, the negative one ($\omega
_{p^{\prime }3}$) of the solutions $\omega _{p^{\prime }}$ in (\ref%
{eq.wpPri23}) is discarded. Consequently, for non-zero bias, there are five
dominant frequencies: $\omega _{p^{\prime }1}=0$, $\omega _{p^{\prime }2}$, $%
\omega _{p1}$, $\omega _{p2}$, $\omega _{p3}$.

For zero bias ($\epsilon =0$), we have $\Gamma (\omega )=\gamma (\omega )$, $%
\Sigma (\omega )=R(\omega )$ and
\begin{equation}
S(\omega )=\frac{\eta ^{2}\Delta ^{2}}{W^{2}}\frac{\Gamma (\omega )}{\left[
\omega -W-R(\omega )\right] ^{2}+\left[ \gamma (\omega )\right] ^{2}}
\end{equation}%
for the non-negative part $\omega \geq 0$. Similarly, the dominant
frequencies can be determined but with only two frequencies $\omega _{p1}$
and $\omega _{p2}$, which are the exactly solvable solutions of $\omega
-W-R(\omega )=0$, as%
\begin{equation}
\omega _{p1,2}=\frac{\Omega +W}{2}\pm \sqrt{\left( \frac{\Omega -W}{2}%
\right) ^{2}+\frac{4g^{2}\eta ^{2}\Delta ^{2}}{(\Omega +W)^{2}}}.
\label{eq.wp12.nobias}
\end{equation}%
It is consistent with Huang's results (see equation (17) in Ref. \cite%
{22huang08}). Compared to unbiased case, the effect of finite bias is three
additional dominant frequencies: $\omega _{p^{\prime }1}=0$, $\omega
_{p^{\prime }2}$, $\omega _{p3}$.

Since the renormalized tunneling in the limit of small HO-environment
coupling is
\begin{equation}
\eta =\exp \left[ -\frac{2g^{2}}{\left( \Omega +W\right) ^{2}}\right] \cong
1,
\end{equation}%
the resonance condition is $\Omega =W_{0}=\sqrt{\Delta ^{2}+\epsilon ^{2}}%
\cong W$. For near resonance $\Omega =W$, (\ref{eq.wp12}) can be simplified
and approximated as%
\begin{equation}
\omega _{p1,2}\cong \Omega \pm \frac{\eta \Delta }{\Omega }g.
\label{eq.wp12.reso}
\end{equation}%
For zero bias, there are only two dominant frequencies $\omega
_{p1,2}=\Omega {\pm }g$ according to (\ref{eq.wp12.nobias}), which is
consistent to the result of the simple exactly solvable Jaynes-Cummings
model \cite{23JC63}. Moreover, for $\Omega =W$ with finite bias, (\ref%
{eq.wpPri23}) can be simplified as%
\begin{equation}
\omega _{p^{\prime }2}=\frac{\sqrt{2}\eta \Delta }{\Omega }g.
\label{eq.wpPri2.reso}
\end{equation}

In the case of finite detunings $\left\vert \Omega -W\right\vert >0$, with
small qubit-HO coupling $g\ll \Delta ,\Omega $, (\ref{eq.wp12}) and (\ref%
{eq.wpPri23}) can be simplified and approximated as
\begin{equation}
\omega _{p1,2}\cong W+\frac{4g^{2}\eta ^{2}\Delta ^{2}}{\left( W-\Omega
\right) (\Omega +W)^{2}}  \label{eq.wp1.offReso}
\end{equation}%
or%
\begin{equation}
\Omega +\frac{4g^{2}\eta ^{2}\Delta ^{2}}{\left( \Omega -W\right) (\Omega
+W)^{2}},  \label{eq.wp2.offReso}
\end{equation}%
(in (\ref{eq.wp1.offReso}) and (\ref{eq.wp2.offReso}), the larger one is $%
\omega _{p1}$ and vice versa) and%
\begin{equation}
\omega _{p^{\prime }2}\cong \left\vert W-\Omega \right\vert +\frac{%
4g^{2}\eta ^{2}\Delta ^{2}}{\left\vert W-\Omega \right\vert (\Omega +W)^{2}}.
\label{eq.wpPri2.offReso}
\end{equation}

It is clear to show the physics of all these dominant frequencies: $\omega
_{p^{\prime }1}=0$ (\ref{eq.wpPri1}) is a relaxation peak, $\omega _{p1}$
and $\omega _{p2}$ are related to the renormalized energy difference of the
qubit in (\ref{eq.wp1.offReso}) and the energy of the HO in (\ref%
{eq.wp2.offReso}), and $\omega _{p^{\prime }2}$ is related to the energy
difference of the qubit and the HO in (\ref{eq.wpPri2.offReso}). Meanwhile, $%
\omega _{p3}$ is related to the summation of the qubit energy $W$ and the HO
energy $\Omega $ as shown in (\ref{eq.wp3}). Therefore, for small qubit-HO
coupling, although $\omega _{p3}$ and $\omega _{p^{\prime }2}$ are not
exactly the summation or the difference between $W$ and $\Omega $, we might
still call $\omega _{p^{\prime }2}$ \textquotedblleft beat
frequency\textquotedblright and $\omega _{p3}$ \textquotedblleft sum
frequency\textquotedblright .

\subsection{Spectrum of the qubit-HO system}

Before exploring the exact spectrum $S(\omega )$ corresponding to the
non-Markovian $P(t)$ in realistic situation, as an alternative view to (\ref%
{Hamiltonian}), we will briefly investigate the energy spectrum of the
equivalent qubit-HO-environment model (\ref{Hamiltonian.equalView}), which
is a physically clearer way. Since the environment generally adds shift and
width to the dominant frequencies, to get a rough idea, the
qubit-HO-environment model without HO-environment coupling ($\Gamma =0$) is
investigated here, and the qubit-HO Hamiltonian reads%
\begin{equation}
H_{\mathrm{q-HO}}=-\frac{\Delta }{2}\sigma _{x}+\frac{\epsilon }{2}\sigma
_{z}+{\Omega }A^{\dagger }A+(A^{\dagger }+A)g\sigma _{z}.
\label{Hamiltonian.equalView.noGamma}
\end{equation}

If the qubit-HO is decoupled ($g=0$), (\ref{Hamiltonian.equalView.noGamma})
is exactly solvable. By applying a unitary matrix to $H_{\mathrm{q-HO}}$, it
can be diagonalized as%
\begin{equation}
H_{\mathrm{q-HO}}=-\frac{W_{0}}{2}\sigma _{z}+{\Omega }A^{\dagger }A.
\label{Hamiltonian.equalView.noGamma.nog}
\end{equation}%
where $W_{0}=\sqrt{\epsilon ^{2}+\Delta ^{2}}$. Thus, the spectrum of the
decoupled qubit-HO without environment is exactly shown, with eigenbasis $%
|n,\pm \rangle $, where $|n\rangle $ denotes the eigenstates of HO with $n$ (%
$n=0,1,\cdots \infty $) phonons and $|\pm \rangle $ denotes the eigenstates
of $\sigma _{z}$ with eigenvalues $\pm 1$ respectively.

If the qubit-HO is switched on ($g\neq 0$), (\ref%
{Hamiltonian.equalView.noGamma}) can be solved with exact numerical
diagonalization, with eigenbasis denoting as $|j\rangle $ ($j=0,1,\cdots
\infty $).

To further explore the instinct of the coupled qubit-HO system, an
analytical deduce beyond RWA is provided as follows. Since (\ref%
{Hamiltonian.equalView.noGamma}) has similar form to (\ref{Hamiltonian})
when removing the summation and the multimode index $k$ and substituting $%
\omega _{k}\rightarrow {\Omega }$, $b_{k}^{\dagger }\rightarrow A^{\dagger }$%
, $b_{k}\rightarrow A$ and $g_{k}/2\rightarrow g$, therefore, we make two
similar unitary transformations $U^{\prime \dag }\exp (S^{\prime })H\exp
(-S^{\prime })U^{\prime }$ to $H$ in (\ref{Hamiltonian.equalView.noGamma})
with generator $S^{\prime }=\left( g/{\Omega }\right) (A^{\dagger }-A)[{%
\Omega }\sigma _{z}/({\Omega }+W^{\prime })-\epsilon /({\Omega }+W^{\prime
})]$ and $U^{\prime }=\left(
\begin{array}{cc}
u^{\prime } & v^{\prime } \\
v^{\prime } & -u^{\prime }%
\end{array}%
\right) $, with $u^{\prime }=\sqrt{\left( 1-\epsilon /W^{\prime }\right) /2}$%
, $v^{\prime }=\sqrt{\left( 1+\epsilon /W^{\prime }\right) /2}$ and $%
W^{\prime }=\sqrt{\epsilon ^{2}+\eta ^{\prime 2}\Delta ^{2}}$, and to the
second order of the qubit-HO coupling $g$ ($O(g^{2})$), it reaches
\begin{eqnarray}
H_{\mathrm{q-HO}} &\cong &-\frac{1}{2}W^{\prime }\sigma _{z}+{\Omega }%
A^{\dagger }A  \nonumber \\
&+&\frac{\epsilon g}{{\Omega }+W^{\prime }}(1-\sigma _{z})(A^{\dagger }+A)
\nonumber \\
&+&\frac{\eta ^{\prime }\Delta g}{{\Omega }+W^{\prime }}\left[ A^{\dagger
}(\sigma _{x}+i\sigma _{y})+A(\sigma _{x}-i\sigma _{y})\right]  \nonumber \\
&-&\frac{g^{2}({\Omega }+2W^{\prime })}{({\Omega }+W^{\prime })^{2}}-\frac{%
g^{2}\epsilon ^{2}}{{\Omega }({\Omega }+W^{\prime })^{2}},
\label{Hamiltonian.equalView.noGamma.withgApprox}
\end{eqnarray}%
where $\eta ^{\prime }=\exp \left[ -2g^{2}/\left( {\Omega }+W^{\prime
}\right) ^{2}\right] $. Note that $(1-\sigma _{z})|+\rangle =0$. If the
value $\left\vert \epsilon g/\left[ \Delta \left( {\Omega }+W^{\prime
}\right) \right] \right\vert \ll 1$, then the term $\epsilon g(1-\sigma
_{z})(A^{\dagger }+A)/\left( {\Omega }+W^{\prime }\right) $ in (\ref%
{Hamiltonian.equalView.noGamma.withgApprox}) can be discarded. Therefore, $%
H_{\mathrm{q-HO}}$ is exactly solvable analytically.

For zero bias ($\epsilon =0$), Ref. \cite{18Hausinger08} has used the Van
Vleck perturbation up to the second order $g$ and solved a Born-Markov
master equation in the system's eigenbasis to get the dynamics $%
P(t)=\sum_{n}p_{nn}(t)+\sum_{n,m (n>m)}p_{nm}(t)$ with the phonon number $n$%
, $m=0,1,\cdots \infty $, and it proposes selection rules for zero bias: $%
p_{nn}(t)$ vanishes for any $n$, and the non-zero $p_{nm}(t)$ only exists
for three cases: $\left\vert n_{even}-m_{even}\right\vert =2$, $\left\vert
n_{odd}-m_{odd}\right\vert =2$, $n_{even}-m_{odd}=3$ or $n_{odd}-m_{even}=1$%
. When substituting $\epsilon =0$ into (\ref%
{Hamiltonian.equalView.noGamma.withgApprox}), the Hamiltonian (\ref%
{Hamiltonian.equalView.noGamma.withgApprox}) is exactly solvable
analytically. Following Ref. \cite{18Hausinger08}, the selection rules can
be deduced similarly. The selection rules show that the transition between
the lowest energy levels $|j\rangle $: $|0\rangle \leftrightarrow |1\rangle $
and $|0\rangle \leftrightarrow |2\rangle $ are allowed, $|1\rangle
\leftrightarrow |2\rangle $ and $|0\rangle \leftrightarrow |3\rangle $ are
forbidden. This offers a second way to explain why there are only two
dominant frequencies for zero bias.

\subsection{Results and discussion}

The Markovian approximation of $P(t)$ is equivalent to approximate the
integration in (\ref{eq.rou21.final}) and (\ref{eq.rou22.final}) by the
residue theorem with single pole at $-2i\gamma _{0}$ and $\omega
_{0}-i\gamma _{0}$, respectively. It leads to
\begin{eqnarray}
&&P(t)=\frac{\eta ^{2}\Delta ^{2}}{W^{2}}\cos (\omega _{0}t)\exp \left(
-\gamma _{0}t\right)  \nonumber \\
&&+\frac{\epsilon }{W}\left[ \left( \frac{\epsilon }{W}+1\right) \exp \left(
-2\gamma _{0}t\right) -1\right] ,  \label{eq.pt.markovian}
\end{eqnarray}%
where $\gamma _{0}=\gamma (W)$ is the Weisskopf-Wigner approximation for the
decay rate and $\omega _{0}=W+\Sigma (W)$ ($\Sigma (W)$ is the level shift).
In the long time limit, the Markovian dynamics is the same as the
non-Markovian one.

In Fig. 3, in the case of zero bias ($\epsilon =0$) with weak coupling ($%
\Delta =\Omega ,$ $g=0.18\Omega ,$ $\alpha =0.004,$ $\Gamma =0.0154$), our
non-Markovian dynamics $P(t)$ and the corresponding spectrum $S(\omega )$ at
zero temperature is compared with the ones by QUAPI \cite{17Thorwart04}, by
VVBM \cite{18Hausinger08} (the numerical results) and by NIBA \cite{21Nesi07}
at low temperature $T=0.1\Delta $. They show good agreement with both $P(t)$
and $S(\omega )$. The reasons of comparing to other results at the low
temperature are: first, corresponding results at zero temperature by other
methods are not found in literature; second, temperature gives a factor $%
\coth \left( \frac{\omega }{2T}\right) $ for each frequency, and in a rough
view $\coth \left( \frac{\omega }{2T}\right) \sim 1$ for typical frequencies
(eg. $\omega =\Delta $) at $T=0.1\Delta $; third, the temperature $%
T=0.1\Delta $ is really low and the comparisons show that their properties
are analogous.

From Fig. 4 to Fig. 6, at finite bias ($\epsilon =-0.5\Delta $, $%
g=0.18\Delta $, $\Gamma =0.0154$), our non-Markovian dynamics $P(t)$ and the
corresponding spectrum $S(\omega )$ at zero temperature is compared with the
numerical results by VVBM \cite{18Hausinger08} at low temperature $%
T=0.1\Delta $ for three different situations: the qubit being at positive
detunings with the HO ($\Omega =1.5\Delta >W_{0}$), on resonance ($\Omega
=W_{0}$), and negative detunings ($\Omega =0.9\Delta <W_{0}$). Roughly, both
the dynamics and the spectrum shows good agreement. Moreover, \textit{our
spectrum presents five dominant frequencies}: one relaxation dip (at $\omega
=0$), one dephasing dip ($\omega _{21}$) and three dephasing peaks ($\omega
_{10}$, $\omega _{20}$, $\omega _{30}$), where the four damping oscillation
frequencies are related to the energy differences of the four lowest energy
levels of the coupled qubit-HO system as shown in the insets and they have
been verified by the exact numerical diagonalization of $H_{\mathrm{q-HO}}$ (%
\ref{Hamiltonian.equalView.noGamma}). Note that the symbols $\omega _{ij}$
denote the dominant frequencies of $S(\omega )$ relating to the energy
difference of the energy levels $|i\rangle $ and $|j\rangle $ of the coupled
qubit-HO system. Meanwhile, the dominant frequencies can also be well
interpreted with the relaxation dip at $\omega =0\Leftrightarrow \omega
_{p^{\prime }1}=0$ in (\ref{eq.wpPri1}), the dephasing peaks at $\omega
_{10}\Leftrightarrow \omega _{p2}$ and $\omega _{20}\Leftrightarrow \omega
_{p1}$ in (\ref{eq.wp12}), $\omega _{30}\Leftrightarrow \omega _{p3}$ in (%
\ref{eq.wp3}), and the dephasing dip at $\omega _{21}\Leftrightarrow \omega
_{p^{\prime }2}$ in (\ref{eq.wpPri23}). Since in Figs. 4 - 6 the qubit-HO
coupling is small ($g=0.18\Delta \ll \Delta ,\Omega $), the expressions for $%
\omega _{p1}$, $\omega _{p2}$ and $\omega _{p^{\prime }2}$ can be written in
simpler approximate forms, as: for on resonance in Fig. 5, $\omega
_{10}\Leftrightarrow \omega _{p2}$ and $\omega _{20}\Leftrightarrow \omega
_{p1}$ in (\ref{eq.wp12.reso}), $\omega _{21}\Leftrightarrow \omega
_{p^{\prime }2}$ in (\ref{eq.wpPri2.reso}); for off-resonance in Fig. 4 and
Fig. 6, $\omega _{10}\Leftrightarrow \omega _{p2}$ and $\omega
_{20}\Leftrightarrow \omega _{p1}$ relating to the renormalized energy
difference of the qubit in (\ref{eq.wp1.offReso}) and relating to the energy
of the HO in (\ref{eq.wp2.offReso}), $\omega _{21}\Leftrightarrow \omega
_{p^{\prime }2}$ relating to the energy difference of the qubit and the HO
in (\ref{eq.wpPri2.offReso}).

In Ref. \cite{18Hausinger08} by VVBM, it presents four dominant frequencies:
one relaxation dip (at $\omega =0$), one dephasing dip ($\omega
_{21}^{\prime }$) and two dephasing peaks ($\omega _{10}^{\prime }$, $\omega
_{20}^{\prime }$), and similar result is also claimed by QUAPI in Ref. \cite%
{17Thorwart04} (see its Fig. 7). In order to distinguish the dominant
frequencies by different methods and/or under different conditions,
analogous symbols $\omega _{ij}^{\prime }$ denote the dominant frequencies
of numerical results by VVBM in Ref. \cite{18Hausinger08} are employed. As
comparison, neither \cite{17Thorwart04} nor \cite{18Hausinger08} presents
the analogous dephasing peak(or dip) at $\omega _{30}$, and to our
knowledge, it is not shown in literature. In all the three figures, the
width and height of the dephasing peak at $\omega _{20}$ matches quite well
with the one at $\omega _{20}^{\prime }$, but our dephasing peak at $\omega
_{10}$ and dephasing dip at $\omega _{21}$ are much higher and sharper,
especially for on resonance in Fig. 5. Meanwhile, the dominant frequencies $%
\omega _{ij}^{\prime }$ are nearly equal to the dominant frequencies $\omega
_{ij}$, but in detail $\omega _{ij}^{\prime }$ are a bit larger than $\omega
_{ij}$. For on resonance in Fig. 5, our dephasing dip at $\omega _{21}$ has
comparable weight with our dephasing peaks at $\omega _{10}$ and $\omega
_{20}$, which is qualitatively different from the ones by VVBM. Besides, we
must admit that the dephasing peaks at $\omega _{30}$ shown in Fig. 4 to
Fig. 6 are really small, which makes it difficult to discover. As a brief
summary, there is a complete new dephasing peak presented in our spectrum
and our dynamics and spectrum shows a good agreement with Ref. \cite%
{18Hausinger08} roughly.

In order to compare our approach with VVBM in Ref. \cite{18Hausinger08}, $g$
are all rather small in Figs. 4 - 6, as well as the corresponding $\alpha $
(all $\alpha <0.005$). Nevertheless, our approach has no direct restriction
in $g$ and it can work with stronger $\alpha $. Therefore, in Figs. 7 - 8,
it presents with larger qubit-HO coupling ($g{=}0.7906\Omega $) and larger
spin-bath coupling ($\alpha =0.01$, $0.05$, $0.1$) for positive detunings ($%
\Omega =2\Delta >W_{0}$) and near resonance ($\Omega =\Delta \sim W_{0}$).
The results for negative detunings are similar to the positive ones and the
figure for negative detunings is not repeated. Likewise, the biased spectrum
presents one relaxation peak (at $\omega =0$) and four dephasing peaks ($%
\omega _{10}$, $\omega _{20}$, $\omega _{21}$, $\omega _{30}$). The dominant
frequencies can be well interpreted with the energy differences of the four
lowest energy levels of the coupled qubit-HO system as above. Similarly,
they can also be well interpreted with the relaxation peak at $\omega
=0\Leftrightarrow \omega _{p^{\prime }1}=0$ in (\ref{eq.wpPri1}), the
dephasing peaks at $\omega _{10}\Leftrightarrow \omega _{p2}$ and $\omega
_{20}\Leftrightarrow \omega _{p1}$ in (\ref{eq.wp12}), $\omega
_{30}\Leftrightarrow \omega _{p3}$ in (\ref{eq.wp3}), and the dephasing dip
at $\omega _{21}\Leftrightarrow \omega _{p^{\prime }2}$ in (\ref{eq.wpPri23}%
). In contrast to the small qubit-HO coupling, the weight of the dephasing
peaks at $\omega _{30}$ shown in Figs. 7 - 8 grows rather larger. For near
resonance in Fig. 8, our dephasing peak at $\omega _{21}$ has comparable
weight with our dephasing peaks at $\omega _{10}$ and $\omega _{20}$ and the
weight of the dephasing peak at $\omega _{30}$ grows much larger than that
for off-resonance in Fig. 7.

Meanwhile, the corresponding Markovian dynamics given by (\ref%
{eq.pt.markovian}) are presented in Fig. 7 to show the long time limit.
Moreover, in Fig. 8 with the same $\Delta =\Omega $, $g=0.7906\Omega $ and $%
\epsilon =0.1\Omega $, the effect with different $\Gamma $ ($=0.002$, $0.01$%
, $0.02$) is shown, as well as different corresponding $\alpha $ ($=0.01$, $%
0.05$, $0.1$). The results show that the distributions of the dominant
frequencies vary little, but with smaller $\Gamma $ or $\alpha $, the
dephasing peaks will be higher and sharper and the dephasing will be
smaller, which is physically reasonable.

The non-Markovian dynamics and the spectrum for zero bias ($\epsilon =0$) in
Figs. 3 - 8 show that the spectrum only presents two dephasing peaks ($%
\omega _{10}^{\prime \prime }$, $\omega _{20}^{\prime \prime }$) for zero
bias, which is consistent with literature results and has been interpreted
in two ways as shown above (with (\ref{eq.wp12.nobias}) or the selection
rules). Similarly, $\omega _{ij}^{\prime \prime }$ are employed to denote
the dominant frequencies for zero bias and they are related to the energy
levels of the unbiased coupled qubit-HO system. Compared to zero bias, the
effect of non-zero bias is shown in Figs. 4 - 8, i.e., three more resonances
in the spectrum appear: the relaxation peak at $\omega =0$, a third
dephasing peak(dip) at $\omega _{21}$ and a fourth dephasing peak at $\omega
_{30}$. Besides, the dominant frequency $\omega _{10}^{\prime \prime }$ for
zero bias is usually smaller than the biased one $\omega _{10}$ and the
dephasing peak at $\omega _{10}^{\prime \prime }$ is usually higher.

For non-zero bias, at $\omega =0$ and $\omega _{21}$, it is clearly shown
two dips for negative bias and two peaks instead for positive bias in Figs.
4 - 8. It can be interpreted as follows: $\omega =0$ is mapped to $\omega
_{p^{\prime }1}=0$ and $\omega _{21}$ is mapped to $\omega _{p^{\prime }2}$;
$\omega _{p^{\prime }1}$ and $\omega _{p^{\prime }2}$ are the solutions $%
\omega _{p^{\prime }}$ of the equation (\ref{eq.omega_pPri}); the equation (%
\ref{eq.omega_pPri}) is from the first term of $S(\omega )$ in (\ref{eq.sw}%
); the sign of the first term of $S(\omega )$ in (\ref{eq.sw}) is the same
with the bias $\epsilon $. Similarly analysis can be done for the remainder
three dominant frequencies. Therefore, there are always peaks at $\omega
_{10}$, $\omega _{20}$ and $\omega _{30}$; while at $\omega =0$ and $\omega
_{21}$, there are peaks for positive bias ($\epsilon >0$), dips for negative
bias ($\epsilon <0$).

As a brief summary to the biased spectrum $S(\omega )$ of the non-Markovian
dynamics $P(t)$: there are five resonances, i.e., the relaxation peak(dip)
at $\omega =0$, one dephasing peak(dip) at $\omega =\omega _{21}$, three
dephasing peaks at $\omega =\omega _{10}$, $\omega _{20}$, $\omega _{30}$,
which are related to the energy differences of the four lowest energy levels
of the coupled qubit-HO system. For the qubit being with the HO at positive
detunings ($\Omega >W_{0}$), the dephasing peak at $\omega =\omega _{10}$
and the relaxation peak(dip) at $\omega =0$ are generally dominant; for
on/near resonance ($\Omega \cong W_{0}$), the dephasing peaks(dip) at $%
\omega =\omega _{10}$, $\omega _{20}$, $\omega _{21}$ are generally
dominant; for negative detunings ($\Omega <W_{0}$), if $g$ is small, the
dephasing peak at $\omega =\omega _{20}$ is generally dominant, otherwise,
the peak at $\omega =\omega _{10}$ is dominant. A rough idea is that: the
dominant frequency(ies) closer to the renormalized energy difference of the
qubit ($W_{0}$) usually contribute(s) more weight.

Fig. 9 shows the dynamics with different $\alpha $ for non-zero bias. As
usual, decay accompanies larger $\alpha $. Our method works well for
sufficient strong spin-bath coupling as long as $\alpha <\alpha _{c}$ (see
Sec. \ref{sec:alpha_C}) beyond weak coupling regime.

The sum rule of the non-Markovian dynamics is checked as shown in Table 1,
and it is exactly satisfied with representative parameters for $\alpha
<\alpha _{c}$.

\section{Susceptibility and coherent-incoherent transition}

\label{sec:alpha_C}

The susceptibility $\chi (\omega )=-G(\omega )$, where $G(\omega )$
(obtained in the Appendix in detail) is the fourier transformation of the
retarded Green's function $G(t)=-i\theta (t)Z^{-1}\mathrm{Tr}\left\{ \exp
(-\beta H)[\sigma _{z}(t),\sigma _{z}]\right\} $, in which $\theta (t)$ is
the unit step function. The imaginary part of $\chi (\omega )$ is $\chi
^{\prime \prime }(\omega )$, as%
\begin{eqnarray}
&&\chi ^{\prime \prime }(\omega )=\frac{\eta ^{2}\Delta ^{2}}{W^{2}}\left\{
\frac{\Gamma (\omega )\theta (\omega )}{[\omega -W-\Sigma (\omega
)]^{2}+\Gamma ^{2}(\omega )}\right.  \nonumber \\
&&-\left. \frac{\Gamma (-\omega )\theta (-\omega )}{[\omega +W+\Sigma
(-\omega )]^{2}+\Gamma ^{2}(-\omega )}\right\} ,  \label{eq.susceptibility}
\end{eqnarray}%
and its real part $\chi ^{\prime }(\omega =0)$ can be obtained by
Kramers-Kronig relation, as
\begin{equation}
\chi ^{\prime }(\omega =0)=\frac{2}{\pi }\int_{0}^{\infty }\frac{\chi
^{\prime \prime }(\omega )}{\omega }\mathrm{d}\omega .
\end{equation}

Our approach can be checked by the Shiba's relation \cite%
{24aShiba90,24bShiba90,25aShiba9698,25bShiba9698}
\begin{equation}
\lim_{\omega \rightarrow 0}\frac{\chi ^{\prime \prime }(\omega )}{J(\omega )}%
={\frac{\pi }{4}}[\chi ^{\prime }(\omega =0)]^{2}.
\end{equation}%
As long as $\alpha <\alpha _{c}$, the Shiba relation is exactly satisfied as
shown in Table. 1 with representative parameters.

The susceptibility $\chi ^{\prime \prime }(\omega )$ is the same as the
second term of $S(\omega )$ in (\ref{eq.sw}) for $\omega \geq 0$ and it is
an odd function of $\omega $. Usually $\chi ^{\prime \prime }(\omega =0)=0$.
While increasing $\alpha $ to a particular value $\alpha _{c}$, a critical
phase happens and $\chi ^{\prime \prime }(\omega =0)=\infty $. Meanwhile, $%
\Gamma (\omega )\propto \omega $ and we have checked that $\left[ \omega
-W-\Sigma (\omega )\right] \propto \omega $ when $\omega \rightarrow 0$.
Thus, the coherent-incoherent transition point \cite{01Leggett87,02Weiss} $%
\alpha _{c}$ is defined as the solution of
\begin{equation}
-W-\Sigma (0)=0.  \label{eq.ac}
\end{equation}

In Fig. 9, non-Markovian dynamics $P(t)$ and the susceptibility $\chi
^{\prime \prime }(\omega )$ are shown with different $\alpha $ ($%
=0.005,0.05,0.1,0.2168$). Meanwhile, the coherent-incoherent transition
point $\alpha _{c}$ ($=0.21683229$) is calculated by (\ref{eq.ac}). $P(t)$
exhibits much abundant oscillation for a weak $\alpha $ ($=0.005$), beating
oscillation for a moderate $\alpha $ ($=0.05$), a badly damping oscillation
for a moderately strong $\alpha $ ($=0.1$), and nearly pure damping for a
sufficient strong $\alpha $ ($=0.2168$) nearly equals to $\alpha _{c}$. In
the inset of Fig. 9, $\chi ^{\prime \prime }(\omega )$ is plotted against $%
\omega $, and the curve shows with three non-zero frequency peaks for all $%
\alpha <\alpha _{c}$. Increasing $\alpha $ from weak ($0.005$) to strong ($%
0.2168$), the peak at the smallest frequency moves rapidly close to $\omega
=0$ and the corresponding peak grows to a great value, and other peaks goes
close to zero. When $\alpha <\alpha _{c}$, $\chi ^{\prime \prime }(\omega
=0)=0$. At $\alpha =\alpha _{c}$, $\chi ^{\prime \prime }(\omega =0)=\infty $%
. Therefore, the particular value $\alpha _{c}$ is the coherent-incoherent
transition point.

In Fig. 10, phase diagrams of the coherent-incoherent transition point $%
\alpha _{c}$ vs. bias $\epsilon $ with different $\Gamma $ are shown. The
area of $\alpha <\alpha _{c}$ is called as the \textquotedblleft coherent
phase\textquotedblright , and the \textquotedblleft incoherent
phase\textquotedblright for $\alpha >\alpha _{c}$. As shown in Fig. 10(a)
for near-resonance $\Delta =\Omega $, the changing curve of $\alpha _{c}$
vs. $\epsilon $ is an Ohmic-alike. $\alpha _{c}$ gradually increases with
increasing bias, and it remarkably increases with increasing $\Gamma $. As
shown in Fig. 10(b) for off-resonance $\Delta =0.5\Omega $, the changing
curve is nontrivial for small $\Gamma =0.075$: one sharp peak exists around $%
\epsilon =0.0655\Omega $ and the curve at the ends behaves Ohmic-alike. The
sharp peak is substituted by a smooth kink for $\Gamma =0.0762$. While for $%
\Gamma =0.08,0.1$, the kink disappears and the whole curves behave
Ohmic-alike. More results show that when decreasing $\Gamma $ (eg. $\Gamma
<0.075$) the sharp peak grows much sharper and higher, and the corresponding
bias of the peak becomes smaller. Under smaller $\Gamma $, a significant
difference of $\alpha _{c}$ between the steep area at finite bias and the
platform area with zero bias might be utilized, eg. reading out the qubit
state.

\section{Conclusions}

\label{sec:conclusion}

We have investigated the biased SBM with a Lorentzian spectral density by a
new analytical approach at zero temperature. An equivalent description of
the system is provided by a biased qubit coupled through a HO to an Ohmic
environment. The starting point is the general SBM Hamiltonian (\ref%
{Hamiltonian}) without RWA. We have applied two unitary transformations to
the Hamiltonian and the non-Markovian master equation within the nontrivial
Born approximation to get an expression for the density operator. With the
density operator, we have provided analytical expressions for the
non-Markovian dynamics $P(t)$ and the corresponding spectrum $S(\omega )$.
Meanwhile, the localized-delocalized transition point $\alpha _{L}$ and the
coherent-incoherent transition point $\alpha _{c}$ are determined, which
have not been provided so far (except $\alpha _{c}$ with zero bias by Ref.
\cite{22huang08}), as well as the analytical ground energy, the renormalized
tunneling factor $\eta $ and the susceptibility $\chi ^{\prime \prime
}(\omega )$. The sum rule and the Shiba relation are carefully checked and
they are exactly satisfied as long as $\alpha <\alpha _{c}$.

The biased dynamics and the corresponding spectrum are key topics in this
paper. Both for biased and unbiased, they have been compared with the
results of the other groups and have shown good agreements. For non-zero
bias, our spectrum presents \textit{five dominant frequencies}: the
relaxation peak(dip) at $\omega =0$, one dephasing peak(dip) at $\omega
=\omega _{21}$, three dephasing peaks at $\omega =\omega _{10}$, $\omega
_{20}$, $\omega _{30}$, and \textit{there is a new effect: an additional
dephasing peak at }$\omega =\omega _{30}$\textit{\ presented in our spectrum}%
. Our approach has no direct restriction on the qubit-HO coupling $g$.
Therefore, it is a good way to investigate the dynamics in the
little-studied strong qubit-HO coupling regime, especially in the static
biased case, which has not been studied yet to our knowledge, as shown in
Figs. 7-9. Moreover, the origin and the meanings of the dominant frequencies
are well studied in two ways: providing analytical expressions for each
dominant frequency in the limit conditions and comparing with the spectrum
of the qubit-HO system. We've also discussed why it is sometimes peak and
sometimes dip, as well as the weight distribution of the peaks(dips) and why
there are only two dominant frequencies for unbiased. Meanwhile, fixing
other parameters, the effect with different $\alpha $ and corresponding $%
\Gamma $ is also shown. The dynamics at the long time limit is given
analytically as $-\epsilon /W$, which is consistent with the Markovian
dynamics.

In summary, we have provided analytical results for interesting physical
quantities without Markovian approximation and our approach works well at
arbitrary detunings: on/off-resonance, with/without bias and for sufficient
strong spin-bath coupling as long as $\alpha <\alpha _{c}$. Admittedly, this
approach is not suitable for very strong spin-bath coupling, e.g. $\alpha
>\alpha _{c}$. Nevertheless, the coherent regime is the most interesting one
in the field of quantum computation and quantum information.

\vskip 0.5cm

{\noindent {\large \textbf{Acknowledgement}}}

This work was supported by the China National Natural Science Foundation
(Grants No.90503007 and No.10734020).

\appendix

\section*{Appendix}

\label{sec:appendix}

\setcounter{equation}{0} \renewcommand{\theequation}{A.\arabic{equation}}

Following the transformation made to $H$ to reach $\tilde{H}$, the retarded
Green's function is
\begin{eqnarray}
&&G(t)=-i\theta (t)Z^{-1}\mathrm{Tr}\left\{ \exp (-\beta H)[\exp (iHt)\sigma
_{z}\exp (-iHt),\sigma _{z}]\right\}  \nonumber \\
&=&-i\theta (t)Z^{-1}\mathrm{Tr}\left\{ \exp (-\beta \tilde{H})\left( \frac{%
\epsilon ^{2}}{W^{2}}[\sigma _{z}(t),\sigma _{z}]+\frac{\eta ^{2}\Delta ^{2}%
}{W^{2}}[\sigma _{x}(t),\sigma _{x}]\right. \right.  \nonumber \\
&&\left. \left. -\frac{\eta \Delta \epsilon }{W^{2}}[\sigma _{z}(t),\sigma
_{x}]-\frac{\eta \Delta \epsilon }{W^{2}}[\sigma _{x}(t),\sigma _{z}]\right)
\right\} .
\end{eqnarray}%
The Fourier transformation of $G(t)$ is denoted as
\begin{eqnarray}
&&G(\omega )=\frac{\epsilon ^{2}}{W^{2}}\left\langle \left\langle \sigma
_{z};\sigma _{z}\right\rangle \right\rangle +\frac{\eta ^{2}\Delta ^{2}}{%
W^{2}}\left\langle \left\langle \sigma _{x};\sigma _{x}\right\rangle
\right\rangle  \nonumber \\
&&-\frac{\eta \Delta \epsilon }{W^{2}}\left\langle \left\langle \sigma
_{z};\sigma _{x}\right\rangle \right\rangle -\frac{\eta \Delta \epsilon }{%
W^{2}}\left\langle \left\langle \sigma _{x};\sigma _{z}\right\rangle
\right\rangle ,
\end{eqnarray}%
where
\[
\left\langle \left\langle A;B\right\rangle \right\rangle =-i\theta (t)Z^{-1}%
\mathrm{Tr}\left\{ \exp (-\beta \tilde{H})[\exp (i\tilde{H}t)A\exp (-i\tilde{%
H}t),B]\right\}
\]%
denotes the retarded Green's function which satisfies the following equation
of motion,
\begin{eqnarray}
&&\omega \left\langle \left\langle A;B\right\rangle \right\rangle
=\left\langle [A,B]\right\rangle +\left\langle \left\langle [A,\tilde{H}%
];B\right\rangle \right\rangle ,  \nonumber \\
&&\left\langle [A,B]\right\rangle =Z^{-1}\mathrm{Tr}\left\{ \exp (-\beta
\tilde{H})[A,B]\right\} .  \nonumber
\end{eqnarray}%
Thus, we can get the following equation chain:%
\begin{eqnarray}
&&\omega \left\langle \left\langle \sigma _{x};\sigma _{x}\right\rangle
\right\rangle =W\left\langle \left\langle i\sigma _{y};\sigma
_{x}\right\rangle \right\rangle  \nonumber \\
&&+\sum_{k}Q_{k}\left\langle \left\langle i\sigma _{y}(b_{k}^{\dag
}+b_{k});\sigma _{x}\right\rangle \right\rangle -\sum_{k}V_{k}\left\langle
\left\langle \sigma _{z}(b_{k}^{\dag }-b_{k});\sigma _{x}\right\rangle
\right\rangle ,
\end{eqnarray}%
\begin{eqnarray}
&&\omega \left\langle \left\langle i\sigma _{y};\sigma _{x}\right\rangle
\right\rangle =2\left\langle \sigma _{z}\right\rangle _{\tilde{H}%
_{0}}+W\left\langle \left\langle \sigma _{x};\sigma _{x}\right\rangle
\right\rangle  \nonumber \\
&&+\sum_{k}Q_{k}\left\langle \left\langle \sigma _{x}(b_{k}^{\dag
}+b_{k});\sigma _{x}\right\rangle \right\rangle +\sum_{k}V_{k}\left\langle
\left\langle \sigma _{z}(b_{k}^{\dag }+b_{k});\sigma _{x}\right\rangle
\right\rangle ,
\end{eqnarray}%
\begin{eqnarray}
&&\omega \left\langle \left\langle \sigma _{x}(b_{k}^{\dag }+b_{k});\sigma
_{x}\right\rangle \right\rangle =-\omega _{k}\left\langle \left\langle
\sigma _{x}(b_{k}^{\dag }-b_{k});\sigma _{x}\right\rangle \right\rangle
\nonumber \\
&&+W\left\langle \left\langle i\sigma _{y}(b_{k}^{\dag }+b_{k});\sigma
_{x}\right\rangle \right\rangle +Q_{k}\left\langle \left\langle i\sigma
_{y};\sigma _{x}\right\rangle \right\rangle ,
\end{eqnarray}%
\begin{eqnarray}
&&\omega \left\langle \left\langle \sigma _{x}(b_{k}^{\dag }-b_{k});\sigma
_{x}\right\rangle \right\rangle =-\omega _{k}\left\langle \left\langle
\sigma _{x}(b_{k}^{\dag }+b_{k});\sigma _{x}\right\rangle \right\rangle
\nonumber \\
&&+W\left\langle \left\langle i\sigma _{y}(b_{k}^{\dag }-b_{k});\sigma
_{x}\right\rangle \right\rangle -Q_{k}\left\langle \left\langle \sigma
_{x};\sigma _{x}\right\rangle \right\rangle ,
\end{eqnarray}%
\begin{eqnarray}
&&\omega \left\langle \left\langle i\sigma _{y}(b_{k}^{\dag }+b_{k});\sigma
_{x}\right\rangle \right\rangle =-\omega _{k}\left\langle \left\langle
i\sigma _{y}(b_{k}^{\dag }-b_{k});\sigma _{x}\right\rangle \right\rangle
\nonumber \\
&&+W\left\langle \left\langle \sigma _{x}(b_{k}^{\dag }+b_{k});\sigma
_{x}\right\rangle \right\rangle +Q_{k}\left\langle \left\langle \sigma
_{x};\sigma _{x}\right\rangle \right\rangle ,
\end{eqnarray}%
\begin{eqnarray}
&&\omega \left\langle \left\langle i\sigma _{y}(b_{k}^{\dag }-b_{k});\sigma
_{x}\right\rangle \right\rangle =-\omega _{k}\left\langle \left\langle
i\sigma _{y}(b_{k}^{\dag }+b_{k});\sigma _{x}\right\rangle \right\rangle
\nonumber \\
&&+W\left\langle \left\langle \sigma _{x}(b_{k}^{\dag }-b_{k});\sigma
_{x}\right\rangle \right\rangle -Q_{k}\left\langle \left\langle i\sigma
_{y};\sigma _{x}\right\rangle \right\rangle ,
\end{eqnarray}%
where $\left\langle \sigma _{z}\right\rangle _{\tilde{H}_{0}}=\left\langle
g_{0}\left\vert \sigma _{z}\right\vert g_{0}\right\rangle =1$. We have
already made the cutoff approximation for the equation chains at the second
order of $g_{k}$. Besides, $\left\langle \left\langle \sigma _{z};\sigma
_{x}\right\rangle \right\rangle =0$, $\left\langle \left\langle \sigma
_{z};\sigma _{z}\right\rangle \right\rangle =0$, and $\left\langle
\left\langle \sigma _{x};\sigma _{z}\right\rangle \right\rangle =0$. So the
solution for $G(\omega )$ is
\begin{eqnarray}
&&G(\omega )=\frac{\eta ^{2}\Delta ^{2}}{W^{2}}\left( \frac{1}{\omega
-W-\sum_{k}V_{k}^{2}/(\omega -\omega _{k})-\sum_{k}Q_{k}^{2}/(\omega
-W-\omega _{k})}\right.  \nonumber \\
&&-\left. \frac{1}{\omega +W-\sum_{k}V_{k}^{2}/(\omega +\omega
_{k})-\sum_{k}Q_{k}^{2}/(\omega +W+\omega _{k})}\right) .
\end{eqnarray}

\newpage

\begin{center}
{\Large \textbf{Figure Captions }}
\end{center}

\vskip 0.5cm

\baselineskip 20pt

\textbf{Fig. 1}~~(color on the web) The renormalized tunneling factor $\eta $
vs. $\alpha $. It shows that $\eta $ goes to zero gradually with smaller $%
\Gamma $ and it goes to zero suddenly with larger $\Gamma $. A slight bias
such as $\epsilon =0.0002\Omega $ can be used to change the curve
effectively. The inset is a magnification part.

\vskip 0.5cm

\textbf{Fig. 2}~~(color on the web) Phase diagram of the
localized-delocalized transition point $\alpha _{L}$ vs. $\epsilon $. It
shows that $\alpha _{L}$ is almost the same with different $\Delta $ and the
curve changes rapidly for larger $\Gamma $ and/or smaller $\epsilon $.
Parameters: near-resonance $\Delta =\Omega $, $\Gamma =0.01$, $0.02$, $0.05$%
, $0.1$; off-resonance $\Delta =0.5\Omega $, $\Gamma =0.01$, $0.02$, $0.05$,
$0.1$.

\vskip 0.5cm

\textbf{Fig. 3}~~(color on the web) For zero bias ($\epsilon =0$),
non-Markovian dynamics $P(t)$ and its corresponding spectrum $S(\omega )$ at
zero temperature show good agreements with results by QUAPI \cite%
{17Thorwart04}, VVBM \cite{18Hausinger08} and NIBA \cite{21Nesi07} at low
temperature $T=0.1\Delta $. Inset (1) is a magnification part and inset (2)
is $S(\omega )$ vs. $\omega $. Parameters: $\Delta =\Omega ,$ $g=0.18\Omega
, $ $\alpha =0.004,$ $\Gamma =0.0154$.

\vskip 0.5cm

\textbf{Fig. 4}~~(color on the web) For off-resonance $\Omega =1.5\Delta $
with $\epsilon =-0.5\Delta $, (a) non-Markovian dynamics $P(t)$ and (b) its
corresponding spectrum $S(\omega )$ at zero temperature shows good agreement
with VVBM \cite{18Hausinger08} at low temperature $T=0.1\Delta $. In detail,
our spectrum presents one relaxation dip (at $\omega =0$), one dephasing dip
($\omega _{21}$) and three dephasing peaks ($\omega _{10}$, $\omega _{20}$, $%
\omega _{30}$). The dominant frequency $\omega _{30}$ is new. Moreover, the
height of our peak at $\omega _{10}$ and dip at $\omega _{21}$ is nearly
twice the ones by VVBM. For zero bias, there are only two dephasing peaks.
Insets: The inset in (a) and insets (1)\symbol{126}(3) in (b) present
magnification parts. Inset (4) in (b) presents the schematic energy levels
of the qubit-HO system with finite bias, where the dashed lines on the left
part for the uncoupled ($g=0$) and the solid lines on the right part are
numerical exact results for the coupled. Our dominant frequencies are
related to the energy differences of the coupled qubit-HO system with
numerical calculation. Parameters: $g=0.18\Delta $, $\Gamma =0.0154$, $%
\alpha {=}0.00177$.

\vskip 0.5cm

\textbf{Fig. 5}~~(color on the web) For on-resonance $\Omega =W_{0}=\sqrt{%
\Delta ^{2}+\epsilon ^{2}}$ with $\epsilon =-0.5\Delta $, (a) non-Markovian
dynamics $P(t)$ and (b) its corresponding spectrum $S(\omega )$ at zero
temperature shows good agreement with VVBM \cite{18Hausinger08} at low
temperature $T=0.1\Delta $. Similarly to Fig. 4, our spectrum presents five
dominant frequencies ($\omega =0$, $\omega _{10}$, $\omega _{20}$, $\omega
_{21}$, $\omega _{30}$) and the dominant frequency $\omega _{30}$ is new.
Our dephasing dip at $\omega _{21}$ is much deeper and sharper than the ones
by VVBM. Our dip at $\omega _{21}$ has comparable weight with our dephasing
peaks at $\omega _{10}$ and $\omega _{20}$, which is qualitatively different
from the ones by VVBM. For zero bias, there are only two dephasing peaks.
Insets: The inset in (a) and inset (1) in (b) presents magnification parts.
Inset (2) in (b) presents the schematic energy levels of the qubit-HO system
with finite bias, where the left dashed lines are for the uncoupled and the
right solid lines are for the coupled. Our dominant frequencies are related
to the energy differences of the coupled qubit-HO system with numerical
calculation. Parameters: $g=0.18\Delta $, $\Gamma =0.0154$, $\alpha {=}%
0.00319$.

\vskip 0.5cm

\textbf{Fig. 6}~~(color on the web) For off-resonance $\Omega =0.9\Delta $
with $\epsilon =-0.5\Delta $, (a) non-Markovian dynamics $P(t)$ and (b) its
corresponding spectrum $S(\omega )$ at zero temperature shows good agreement
with VVBM \cite{18Hausinger08} at low temperature $T=0.1\Delta $. Similarly
to Fig. 4, our spectrum presents five dominant frequencies ($\omega =0$, $%
\omega _{10}$, $\omega _{20}$, $\omega _{21}$, $\omega _{30}$) and the
dominant frequency $\omega _{30}$ is new. Moreover, our peaks (or dips) are
a bit sharper and higher than the ones by VVBM. For zero bias, there are
only two dephasing peaks. Insets: The inset in (a) and inset (1) in (b)
presents magnification parts. Inset (2) in (b) presents the schematic energy
levels of the qubit-HO system with finite bias, where the left dashed lines
are for the uncoupled and the right solid lines are for the coupled. Our
dominant frequencies are related to the energy differences of the coupled
qubit-HO system with numerical calculation. Parameters: $g=0.18\Delta $, $%
\Gamma =0.0154$, $\alpha {=}0.00493$.

\vskip 0.5cm

\textbf{Fig. 7}~~(color on the web) For off-resonance $\Delta =0.5\Omega $
with larger qubit-HO coupling $g{=}0.7906\Omega $, (a) non-Markovian
dynamics $P(t)$ and (b) its corresponding spectrum $S(\omega )$ at zero
temperature is shown with bias $\epsilon =0.1\Omega $ and with zero bias,
while the Markovian ones in (a) has shown the long time limit. Similarly,
the biased spectrum presents four dephasing peaks ($\omega _{10}$, $\omega
_{20}$, $\omega _{21}$, $\omega _{30}$) and one relaxation peak ($\omega =0$%
), while the unbiased one only presents two dephasing peaks. Insets: Inset
(1) in (b) presents a magnification part. Inset (2) in (b) presents the
schematic energy levels of the qubit-HO system with finite bias, where the
left dashed lines are for the uncoupled and the right solid lines are for
the coupled. Our dominant frequencies are related to the energy differences
of the coupled qubit-HO system with numerical calculation. Parameters: $%
\alpha =0.01$, $\Gamma =0.002$.

\vskip 0.5cm

\textbf{Fig. 8}~~(color on the web) For near-resonance $\Delta =\Omega $
with larger qubit-HO coupling $g{=}0.7906\Omega $, (a) non-Markovian
dynamics $P(t)$ and (b) its corresponding spectrum $S(\omega )$ at zero
temperature is compared between the biased $\epsilon =0.1\Omega $ and the
unbiased at $\alpha =0.05$, $\Gamma =0.01$. Similarly, the biased spectrum
presents four dephasing peaks ($\omega _{10}$, $\omega _{20}$, $\omega _{21}$%
, $\omega _{30}$) and one relaxation peak ($\omega =0$), while the unbiased
one only presents two dephasing peaks. Besides, with the same $\Delta
=\Omega $, $g=0.7906\Omega $ and $\epsilon =0.1\Omega $, the effect with
different $\Gamma $ or $\alpha $ (red dot line: $\alpha =0.01$, $\Gamma
=0.002$ and violet dash line: $\alpha =0.1$, $\Gamma =0.02$) is shown: the
distributions of the dominant frequencies vary little, but with smaller $%
\Gamma $ or $\alpha $, the dephasing peaks will be higher and sharper and
the dephasing will be smaller. Insets: Inset in (b) presents the schematic
energy levels of the qubit-HO system with finite bias where the left dashed
lines are for the uncoupled and the right solid lines are for the coupled.
Our dominant frequencies are related to the energy differences of the
coupled qubit-HO system with numerical calculation.

\vskip0.5cm

\textbf{Fig. 9}~~(color on the web) Non-Markovian dynamics $P(t)$ for the
near resonance case $\Delta =\Omega $ with $\epsilon =0.1\Omega $ and $%
\Gamma =0.01$ for different spin-bath coupling $\alpha =0.005$, $0.05$, $0.1$%
, $0.2168$ (corresponding qubit-HO coupling $g/\Omega =0.25$, $0.7906$, $%
1.1180$, $1.6462$) with all $\alpha <\alpha _{c}=0.21683229$. It shows that
when increasing $\alpha $ from weak to strong, the dynamics goes from
abundant oscillation to nearly pure damping. Insets: The inset is the
susceptibility $\chi ^{\prime \prime }(\omega )$ vs. $\omega $ and it shows
that when increasing $\alpha $ close to $\alpha _{c}$ the highest peak goes
close to infinity near $\omega =0$.

\vskip0.5cm

\textbf{Fig. 10}~~Phase diagram of the coherent-incoherent transition point $%
\alpha _{c}$ vs. bias $\epsilon $, for (a) near-resonance $\Delta =\Omega $
with different $\Gamma =0.05$, $0.08$, $0.1$; (b) off-resonance $\Delta
=0.5\Omega $ with different $\Gamma =0.075$, $0.0762$, $0.08$, $0.1$. A kink
appears when $\Gamma $ is small for off-resonance, otherwise $\alpha _{c}$
gradually increases with $\epsilon $.

\newpage

\begin{center}
{\Large \textbf{Tables }}
\end{center}

\textbf{Table 1}~~The sum rule and the Shiba relation are checked with
representative parameters. Here $R\equiv \lim_{\omega \rightarrow 0}F(\omega
)/\frac{\pi }{4}\left[ \chi ^{\prime }(0)\right] ^{2}$, where $F(\omega
)=\chi ^{\prime \prime }(\omega )/J(\omega )$.

\begin{table}[th]
\centering\setlength\tabcolsep{3pt}
\begin{tabular}{cccccccc}
\hline
$\alpha $ & $\Delta /\Omega $ & $\Gamma $ & $\epsilon /\Omega $ & $\chi
^{\prime }(0)$ & $\lim_{\omega \rightarrow 0}F(\omega )$ & $R$ & $P(t=0)$ \\
\hline
&  &  &  &  &  &  &  \\
0.1 & 0.1 & 0.05 & 0.1 & 5.897376962 & 27.31540594 & 1 & 1 \\
0.1 & 0.2 & 0.05 & 0.1 & 10.16150527 & 81.09722154 & 1 & 1 \\
0.1 & 0.5 & 0.01 & 0 & 77.38519588 & 4703.332194 & 1 & 1 \\
0.3 & 0.5 & 0.01 & 0.01 & 0.05845719399 & 0.002683896792 & 1 & 1 \\
0.1 & 0.5 & 0.1 & 0.5 & 1.558179498 & 1.906886539 & 1 & 1 \\
0.2 & 0.5 & 0.2 & 0.1 & 5.784656542 & 26.28119072 & 1 & 1 \\
0.1 & 1 & 0.01 & 0 & 49.76240845 & 1944.879348 & 1 & 1 \\
0.3 & 1 & 0.01 & 0.01 & 0.233784401 & 0.04292605141 & 1 & 1 \\
0.1 & 1 & 0.01 & 0.1 & 36.23722191 & 1031.3348 & 1 & 1 \\
0.21 & 1 & 0.01 & 0.1 & 7.819123532 & 48.01821704 & 0.9999999999 & 1 \\
0.216 & 1 & 0.01 & 0.1 & 49.36762968 & 1914.143282 & 1.000000025 & 1 \\
0.2168 & 1 & 0.01 & 0.1 & 1228.943521 & 1186131.33 & 0.9999517554 &
0.9999999947 \\
0.2 & 1 & 0.05 & 0.01 & 7.857533292 & 48.49113404 & 1 & 1 \\
0.5 & 1 & 0.2 & 0.1 & 5.312073901 & 22.1624668 & 1 & 1 \\
0.1 & 1.1 & 0.05 & 0.1 & 2.920164619 & 6.697373981 & 1 & 1 \\
0.2 & 1.2 & 0.01 & 0.1 & 7.494084145 & 44.10897966 & 1 & 1 \\
0.1 & 1.5 & 0.05 & 0.1 & 2.003650314 & 3.15307092 & 1 & 1 \\ \hline
\end{tabular}%
\end{table}

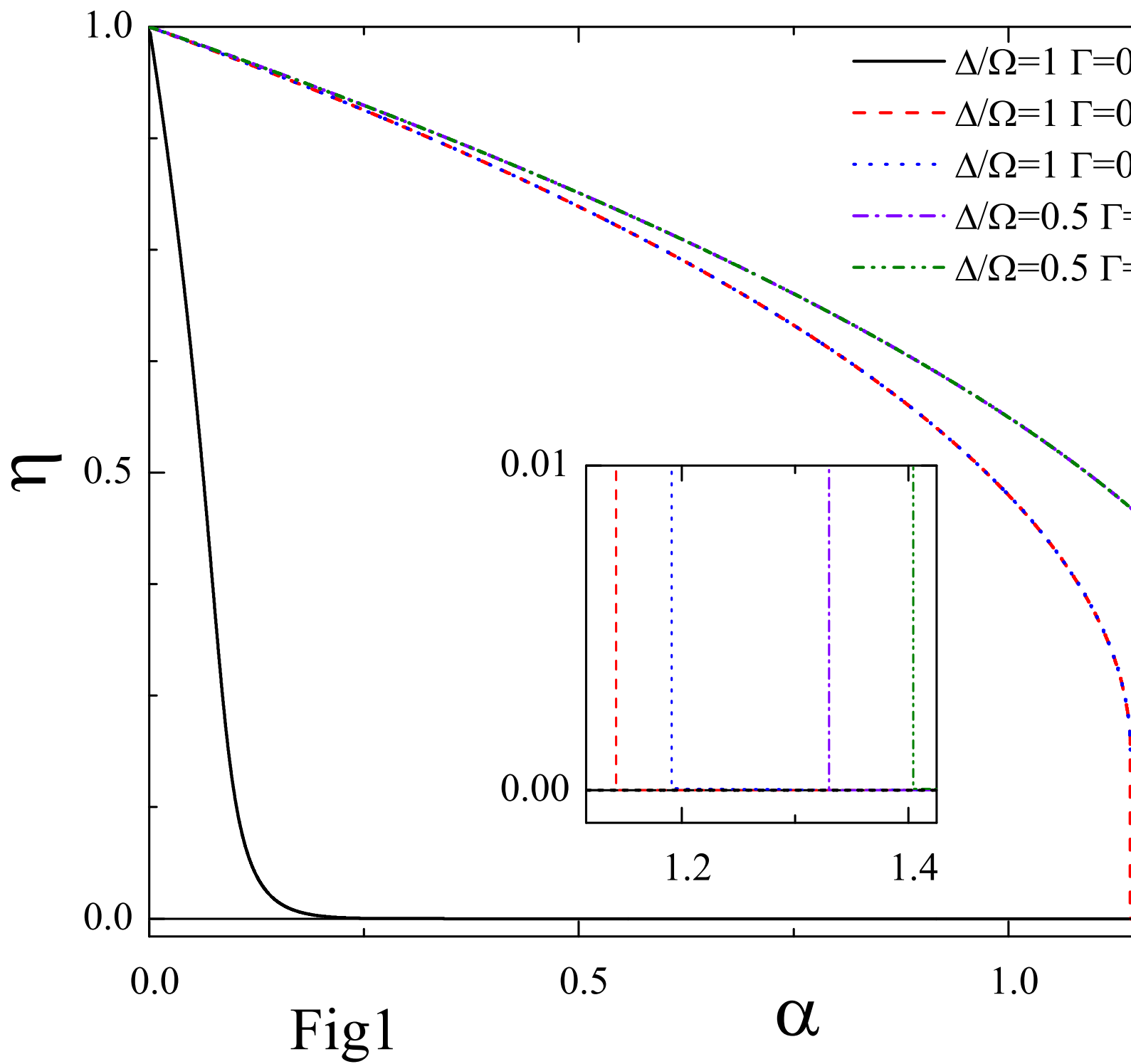
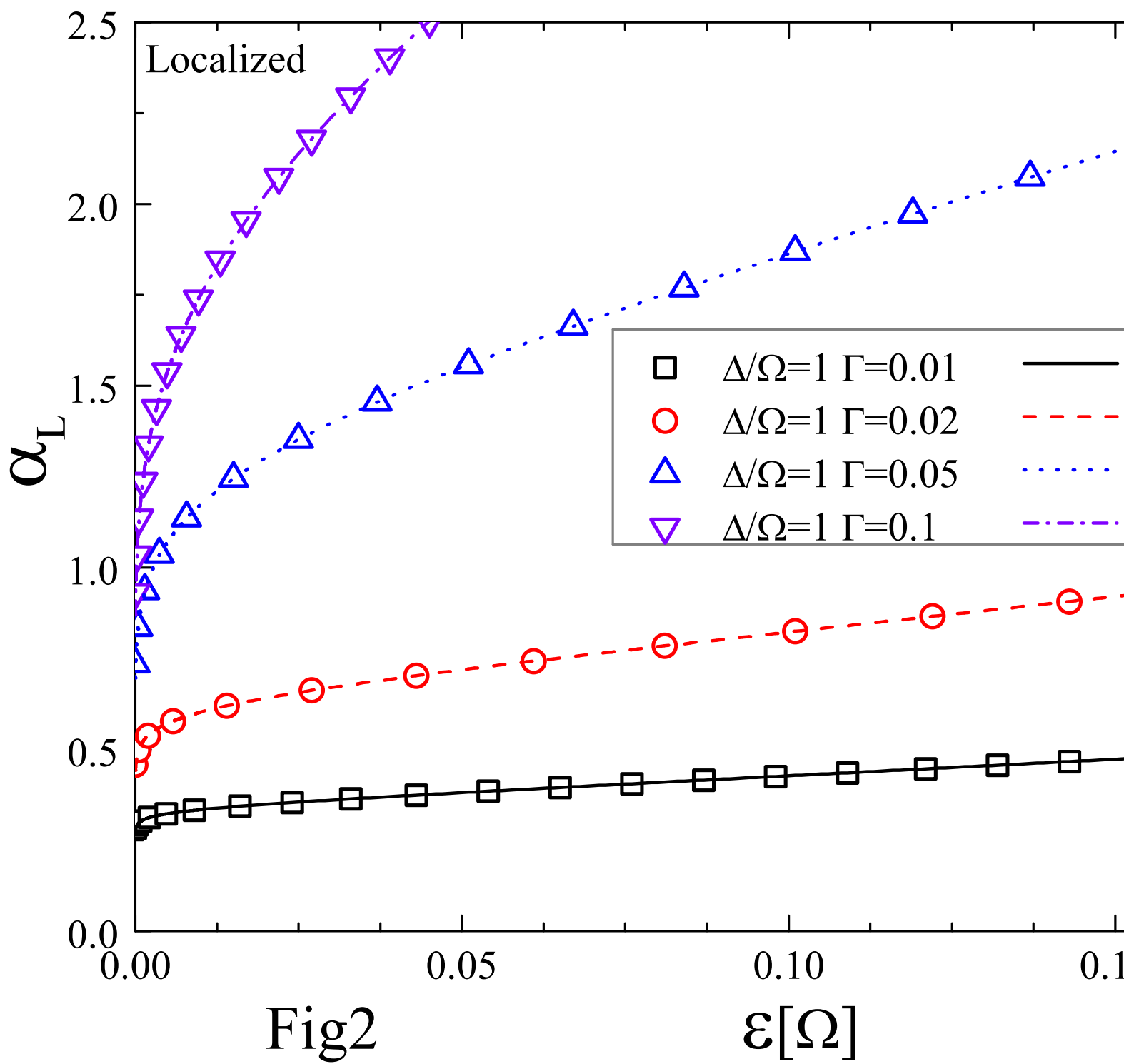
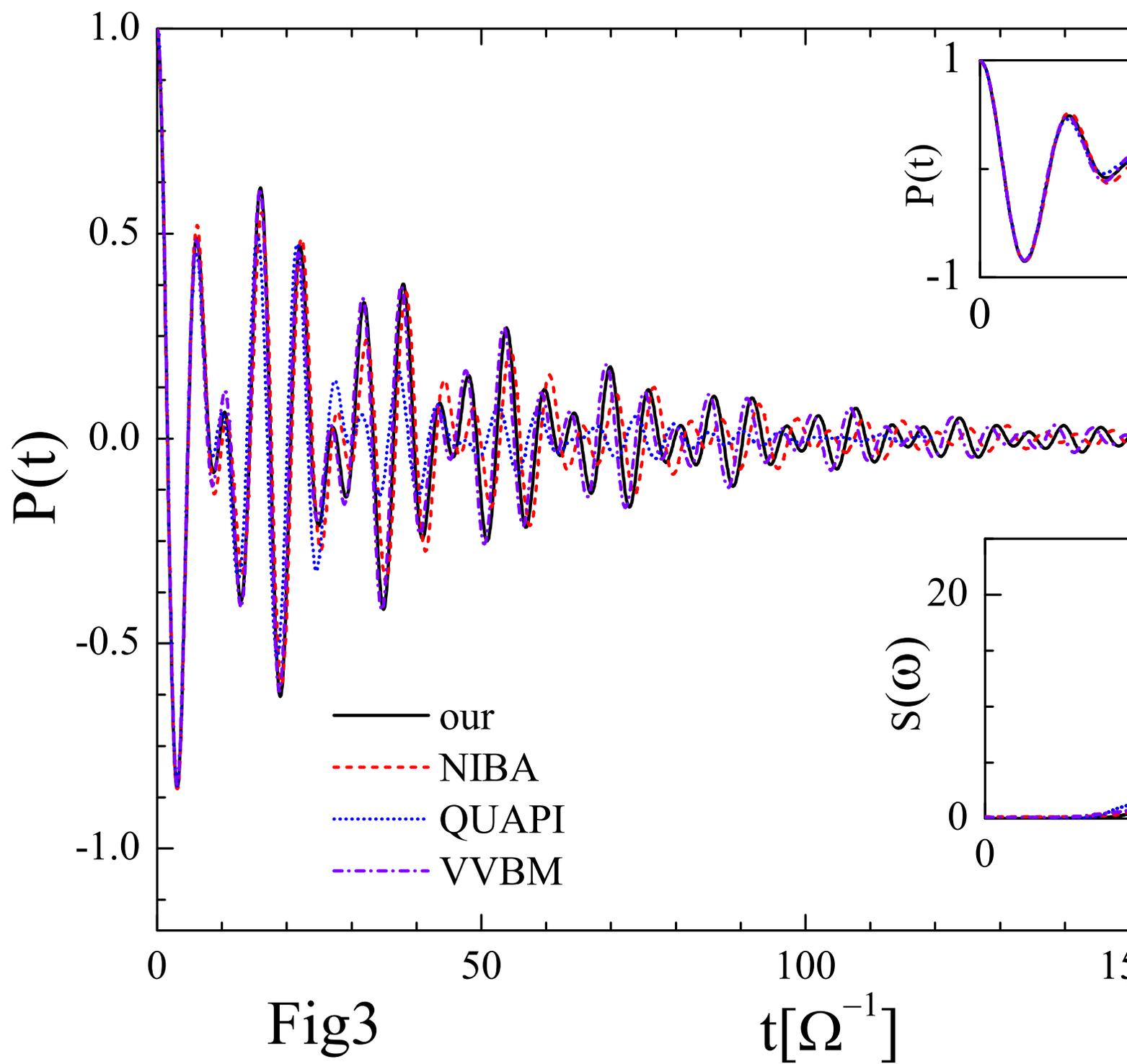
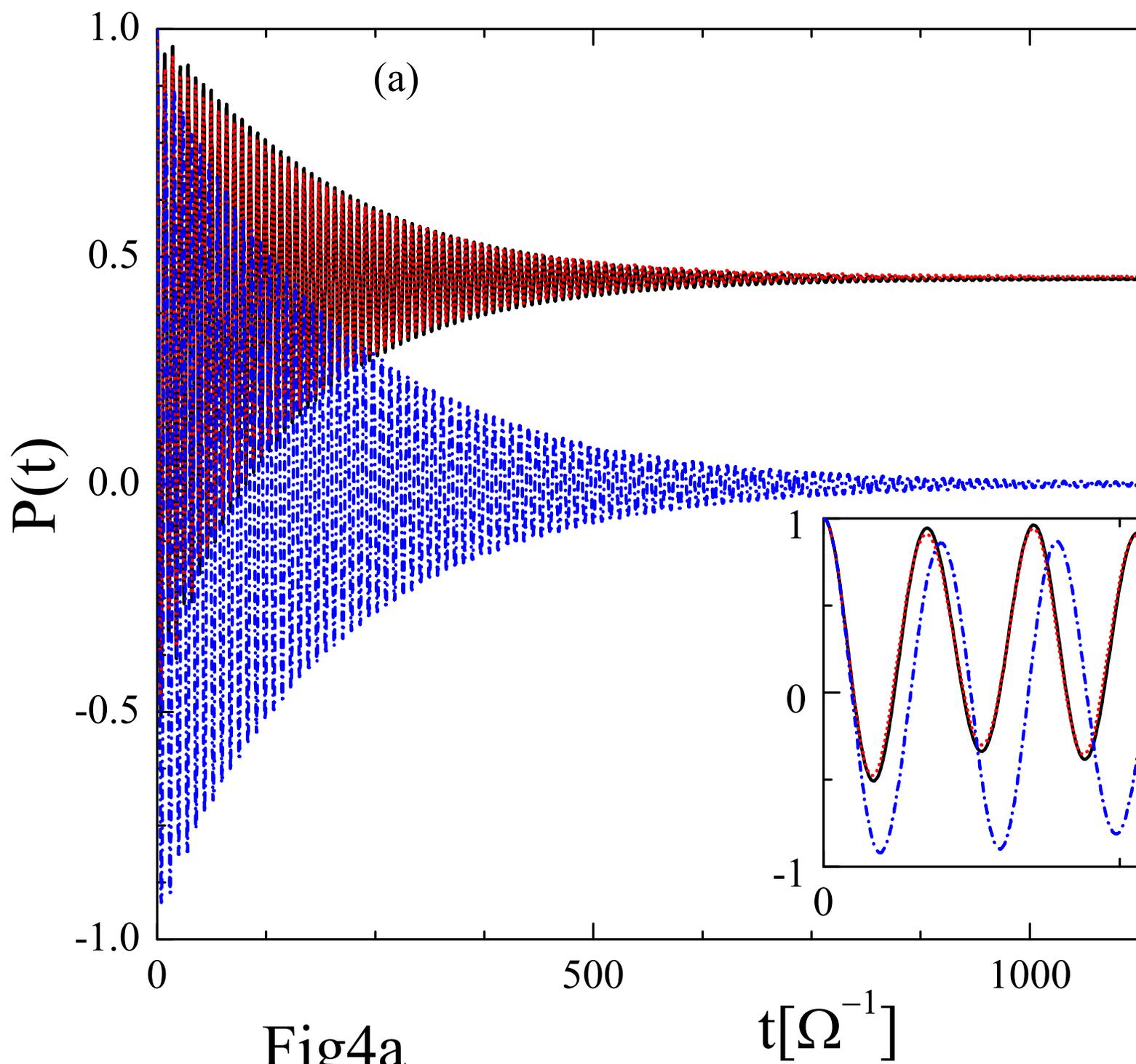
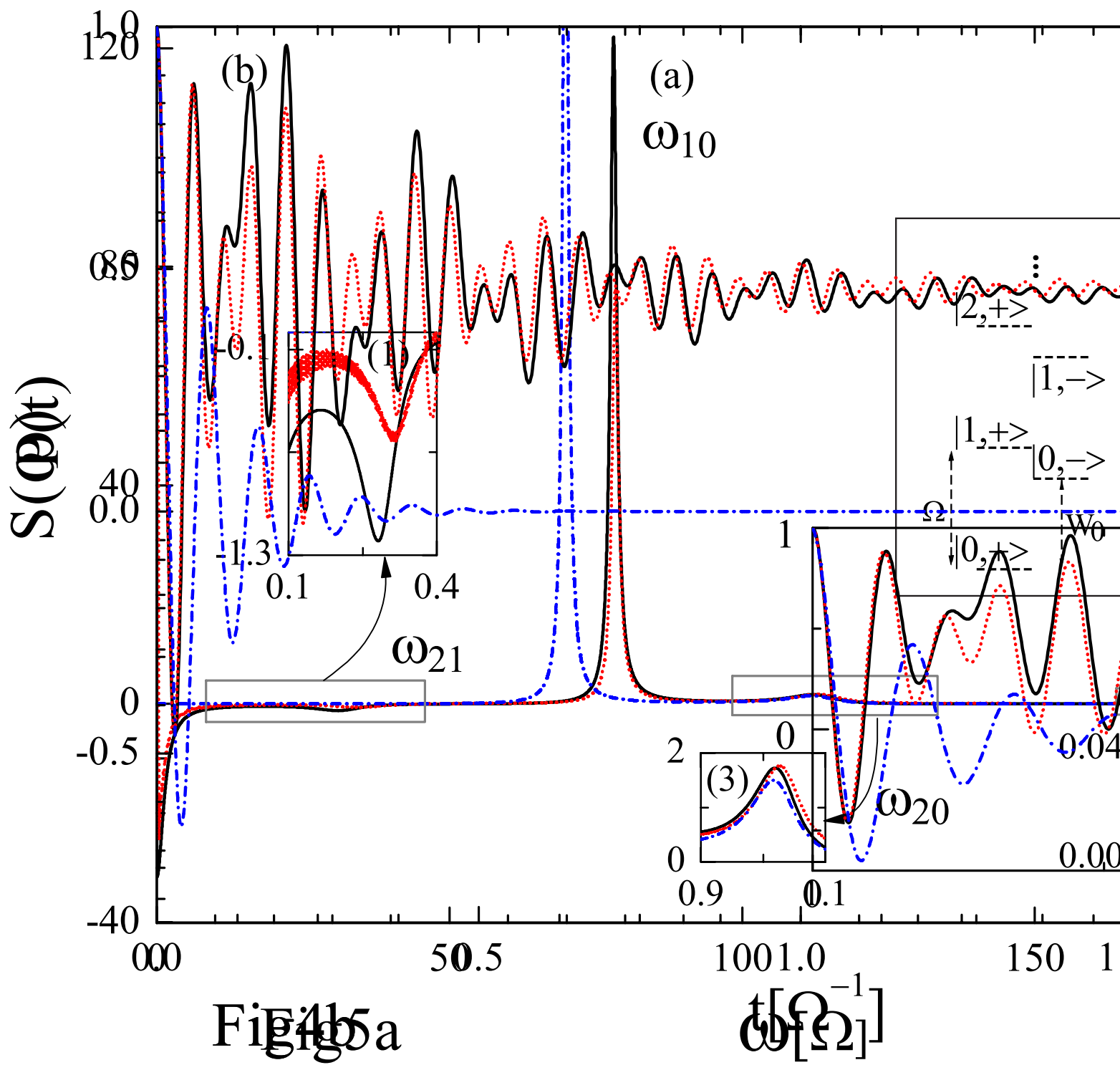
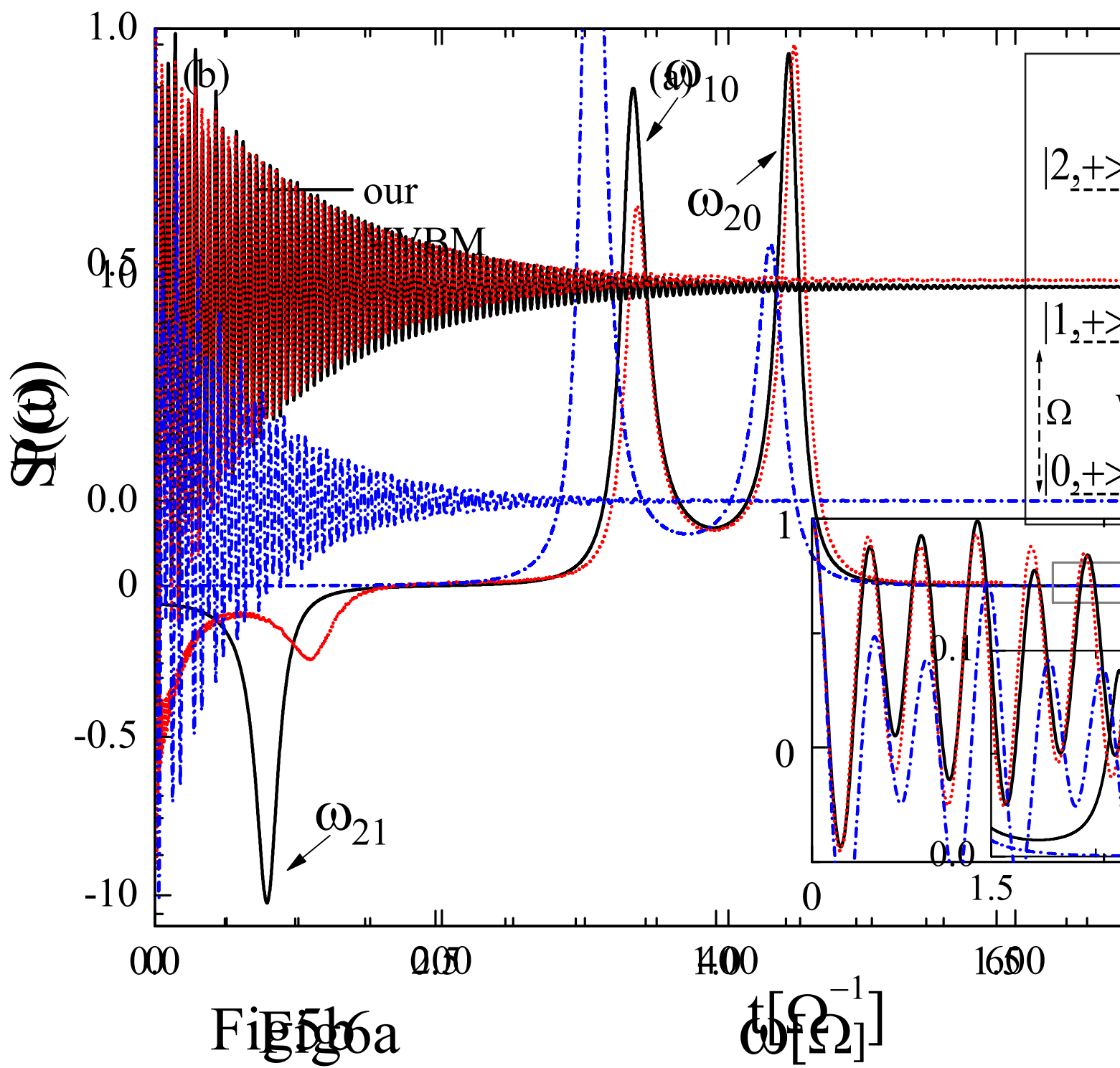
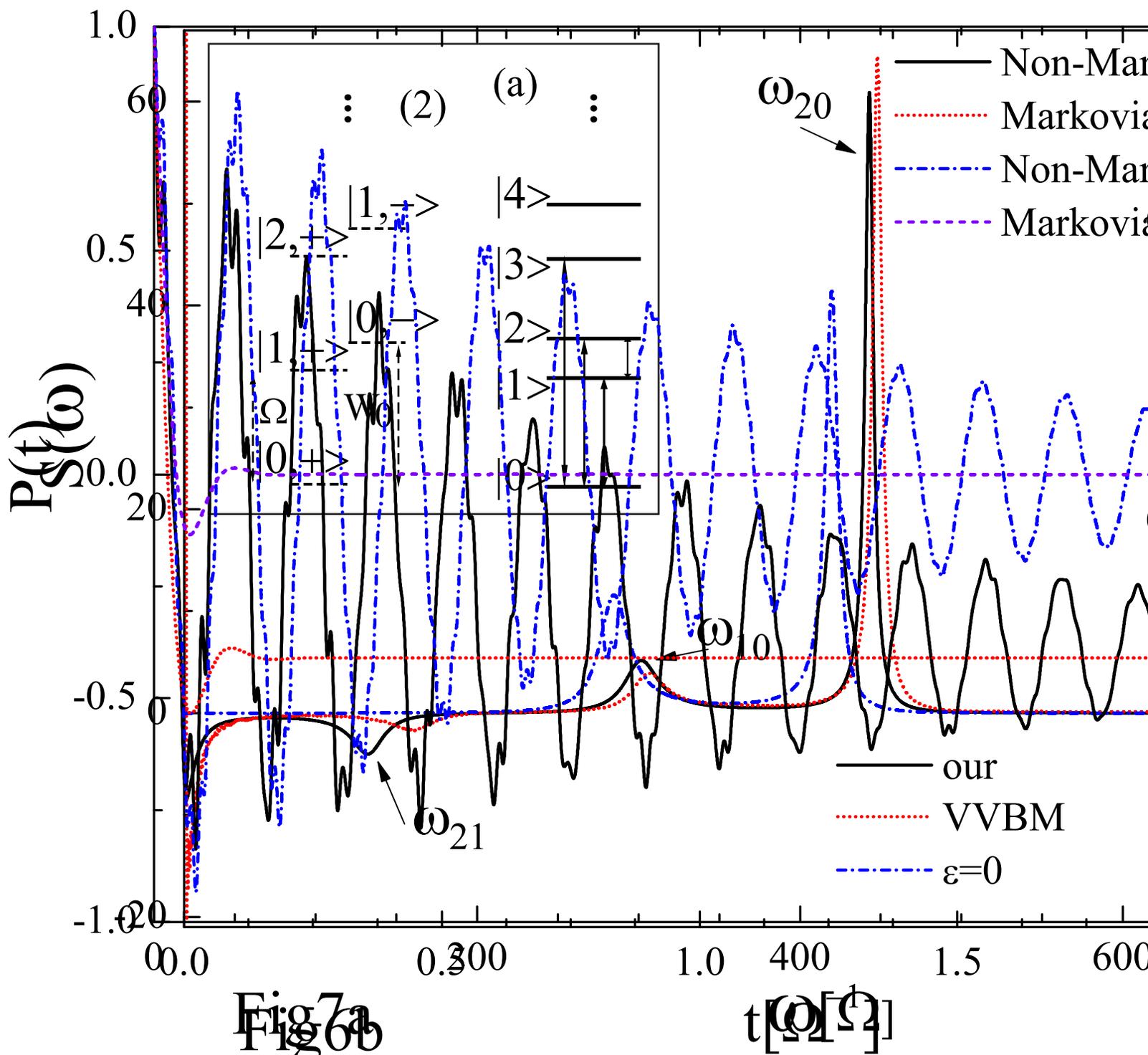
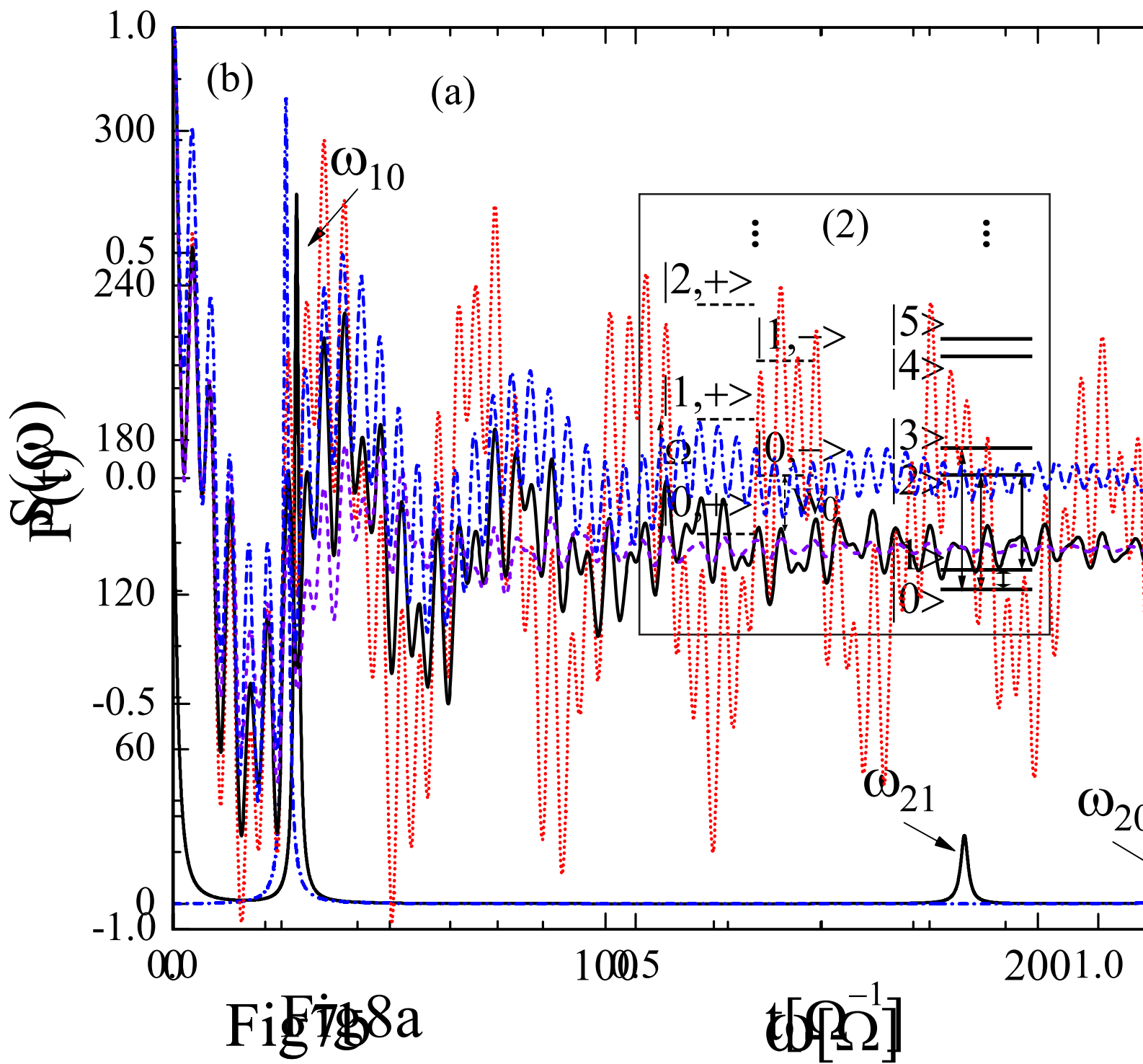
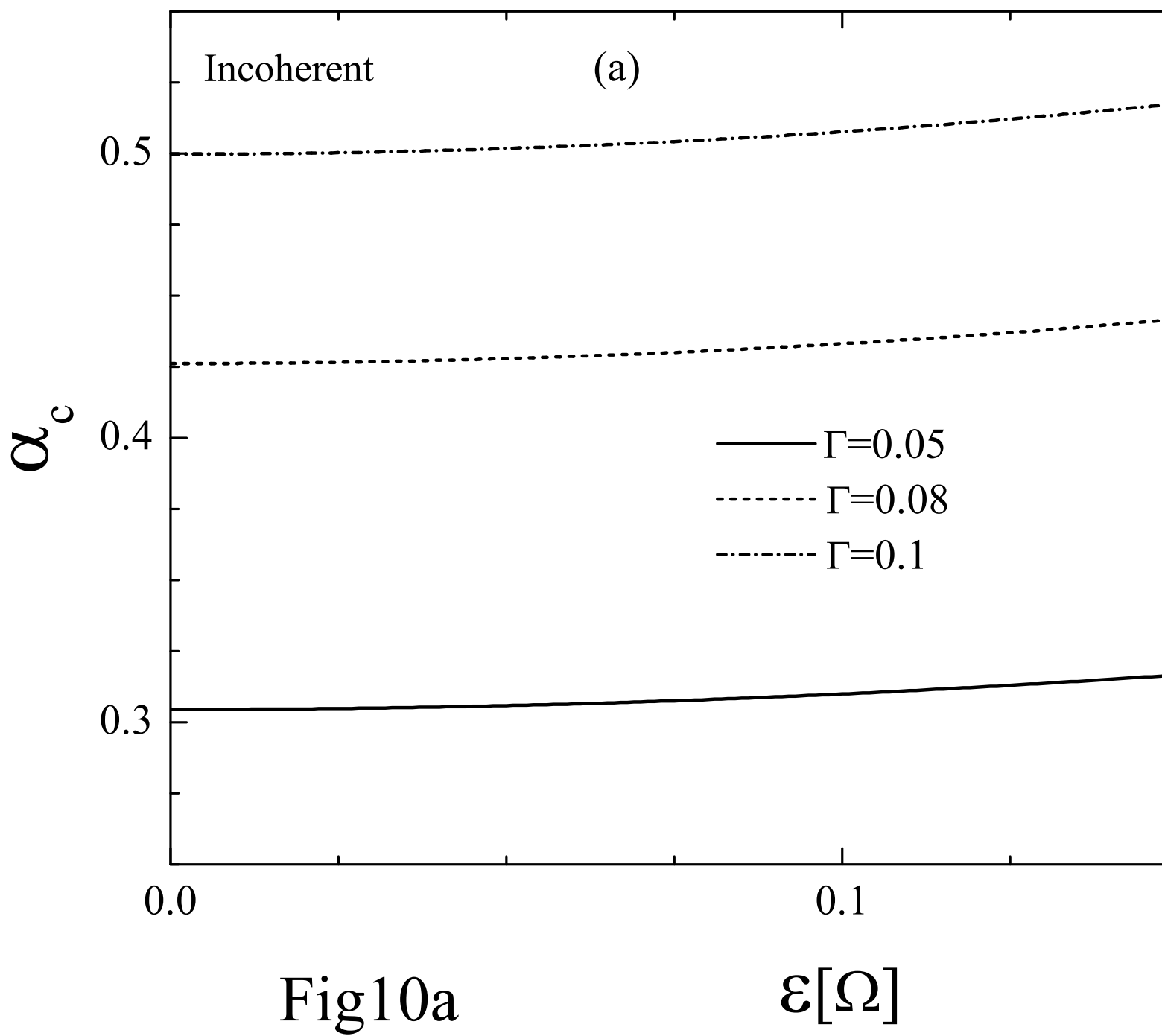
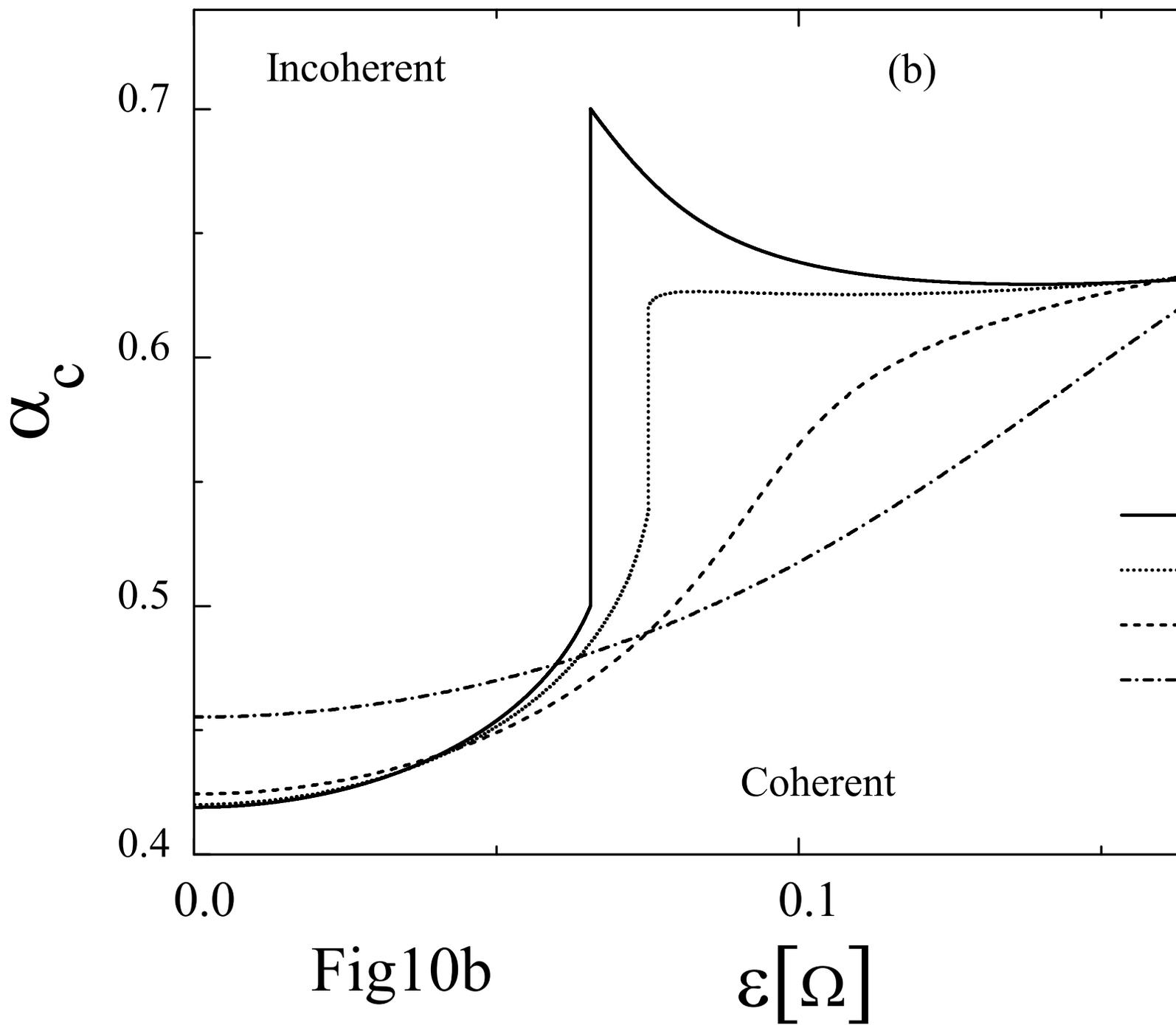
\end{document}